\documentclass{emulateapj}
\usepackage{graphicx}
\usepackage{epsfig}
\usepackage{natbib}
\usepackage{multirow}
\usepackage{float}
\usepackage[caption=false]{subfig}

\newcommand{\eg}{{\rm e.g.,\ }}

\newcommand{\ie}{{\rm i.e.,\ }}
\newcommand{\cf}{{\rm cf.\ }}

\newcommand{\km}{{\rm\thinspace km}}

\newcommand{\s}{{\rm\thinspace s}}

\newcommand{\erg}{{\rm\thinspace ergs}}

\newcommand{\ergps}{\hbox{$\erg\s^{-1}\,$}}

\newcommand{\kmps}{\hbox{$\km\s^{-1}\,$}}

\newcommand{\ha}{H$\alpha$}

\newcommand{\hi}{H\thinspace{\sc i}}
\newcommand{\hii}{H\thinspace{\sc ii}}
\newcommand{\htwo}{H$_2$}
\newcommand{\nii}{[N\thinspace{\sc ii}]}

\newcommand{\mhi}{$M_{\rm HI}$}
\newcommand{\gf}{\mhi/$M_*$}

\newcommand{\Msol}{\hbox{\thinspace M$_{\sun}$}}
\newcommand{\Zsol}{\hbox{\thinspace Z$_{\sun}$}}

\slugcomment{Submitted version; draft version \today}

\shorttitle{\hi~and Star Formation in ALFALFA H$\alpha$}
\shortauthors{Jaskot, Oey, Salzer, Van Sistine, Bell, Haynes}

\begin{document}

\title{From \hi\ to Stars: \hi\ Depletion in Starbursts and Star-Forming Galaxies in the ALFALFA~\ha\ Survey}

\author{A. E. Jaskot, \altaffilmark{1}$^,$\altaffilmark{2}
	M. S. Oey, \altaffilmark{2}
	J. J. Salzer, \altaffilmark{3}$^,$\altaffilmark{4}
	A. Van Sistine, \altaffilmark{3}$^,$\altaffilmark{4}
	E. F. Bell, \altaffilmark{2}
	M. P. Haynes, \altaffilmark{5}
	}
\altaffiltext{1}{Department of Astronomy, Smith College, Northampton, MA 01063, USA.}
\altaffiltext{2}{Department of Astronomy, University of Michigan, Ann Arbor, MI 48109, USA.}
\altaffiltext{3}{Department of Astronomy, Indiana University, Bloomington, IN 47405, USA.}
\altaffiltext{4}{Visiting Astronomer, Kitt Peak National Observatory, National Optical Astronomy Observatory, which is operated by the Association of Universities for Research in Astronomy (AURA) under cooperative agreement with the National Science Foundation.}
\altaffiltext{5}{Center for Radiophysics and Space Research, Space Sciences Building, Cornell University, Ithaca, NY 14853, USA.}

\begin{abstract}
\hi\ in galaxies traces the fuel for future star formation and reveals the effects of feedback on neutral gas. Using a statistically uniform, \hi-selected sample of 565 galaxies from the ALFALFA~H$\alpha$\ survey, we explore \hi\ properties as a function of star formation activity. ALFALFA~H$\alpha$\ provides $R$-band and H$\alpha$\ imaging for a volume-limited subset of the 21-cm ALFALFA survey. We identify eight starbursts based on H$\alpha$\ equivalent width and six with enhanced star formation relative to the main sequence. Both starbursts and non-starbursts have similar \hi\ to stellar mass ratios (\mhi/$M_*$), which suggests that feedback is not depleting the starbursts' \hi. Consequently, the starbursts do have shorter \hi\ depletion times ($t_{\rm dep}$), implying more efficient \hi-to-\htwo\ conversion. While major mergers likely drive this enhanced efficiency in some starbursts, the lowest mass starbursts may experience periodic bursts, consistent with enhanced scatter in $t_{\rm dep}$\ at low $M_*$. Two starbursts appear to be pre-coalescence mergers; their elevated \mhi/$M_*$\ suggest that \hi-to-\htwo\ conversion is still ongoing at this stage. By comparing with the GASS sample, we find that $t_{\rm dep}$\ anti-correlates with stellar surface density for disks, while spheroids show no such trend. Among early-type galaxies, $t_{\rm dep}$\ does not correlate with bulge-to-disk ratio; instead, the gas distribution may determine the star formation efficiency. Finally, the weak connection between galaxies' specific star formation rates and \mhi/$M_*$\ contrasts with the well-known correlation between \mhi/$M_*$\ and color. We show that dust extinction can explain the \hi-color trend, which may arise from the relationship between $M_*$, \mhi, and metallicity.

\end{abstract}

\section{Introduction}
Gas inflows and outflows drive galaxy evolution by controlling the raw material from which stars form. The star formation history of the Universe may reflect the history of gas accretion onto dark matter halos \citep[\eg][]{keres05,prochaska09}, and observed galaxy scaling relations may likewise trace the history of gas flows. For instance, the observed relations between galaxy stellar masses, star formation rates (SFRs), and metallicities may stem from variations in the efficiency with which galaxies accrete and expel gas \citep[\eg][]{dalcanton07,mannucci10,dave11a,dave11b,lilly13}.

The connection between gas content and star formation is a well-established result. Individual stars form from dense cores within molecular clouds \citep[\eg][]{myers83,motte98}. On kiloparsec scales, a galaxy's SFR surface density, $\Sigma_{\rm SFR}$, increases with the \hi+\htwo~gas surface density, $\Sigma_{\rm HI+H_2}$, as parameterized by the Kennicutt-Schmidt Law \citep{kennicutt98}. More recent work shows that $\Sigma_{\rm H_2}$, rather than $\Sigma_{\rm HI}$, drives the Kennicutt-Schmidt law, even in \hi-dominated regimes \citep{schruba11}. Above a threshold density of $\sim$10 \Msol/pc$^2$, \hi~``saturates"; this column density is sufficient to shield molecular gas from photodissociation. Most gas above this threshold is molecular \citep[\eg][]{wong02,bigiel08}, resulting in no trend between SFR and \hi~in this density regime. Below this threshold, however, $\Sigma_{\rm SFR}$~and $\Sigma_{\rm HI}$~correlate, albeit with a large scatter, due to the relation between $\Sigma_{\rm HI}$~and $\Sigma_{\rm H_2}$ \citep{schruba11}. 

Compared to the link between \htwo~content and SFR, the relationship between galaxies' \hi~content and star formation is less straightforward. Galaxies' \hi~typically extends to much larger radii than the stellar distribution \citep[\eg][]{broeils97} and may constitute a gas reservoir for fueling future star formation. Accretion of gas from the intergalactic medium may replenish this reservoir, and gas flows may bring \hi~inward, leading to star formation in the inner regions of galaxies. \citet{prochaska09}~suggest that galaxies' \hi~disks exist at a constant, unstable density, with any subsequent accretion leading to the creation of stars and resulting in an SFR that traces the accretion rate. In massive galaxies, the total \hi~gas fraction correlates with signs of recent accretion, such as an outer metallicity drop, and this accretion appears to power star formation throughout the galaxy disk \citep{moran12}. The \hi~in low-mass galaxies, on the other hand, may not signify recent accretion. Low-mass galaxies tend to be more gas-rich than high-mass galaxies, and \hi~constitutes the dominant component of their interstellar medium (ISM). The long gas consumption times in such systems may indicate that star formation proceeds inefficiently. Alternatively, \citet{kannappan13}~argue that star formation cannot keep pace with the rate of gas accretion, causing \hi\ to accumulate. The relationship between \hi~and star formation may also change in dwarf galaxies due to their lower metallicities. At low metallicities and consequently, low dust content, the formation of a given $\Sigma_{\rm H_2}$~requires a higher \hi~column density \citep{krumholz09}. As a result, the relationship between \hi~density and \htwo~formation differs for low-metallicity galaxies, with galaxies such as the Small Magellanic Cloud exhibiting a higher threshold density for \hi~saturation \citep{bolatto11}.

The advent of large \hi~surveys capable of resolving individual galaxies has clarified the connection between \hi~mass and galaxy properties. In particular, the Arecibo Legacy Fast ALFA (ALFALFA) survey is a blind 21-cm survey, which covers 7000 deg$^2$~and has detected $\sim$30,000 galaxies out to $z=0.06$ \citep{giovanelli05a,giovanelli05b,haynes11}. In conjunction with ALFALFA, the {\it GALEX} Arecibo SDSS (GASS; \citealt{catinella10}) and H$\alpha$3 \citep{gavazzi12} surveys have investigated \hi\ and star formation in different galaxy regimes. The GASS survey examines the \hi~content of massive galaxies ($M_* > 10^{10}$ \Msol) using additional Arecibo observations for galaxies undetected in ALFALFA. H$\alpha$3 studies the effect of environment on \hi\ content and star formation in ALFALFA galaxies within the Local Supercluster. The ALFALFA survey and related surveys have established scaling relations between \hi~gas fraction and galaxy stellar mass, stellar surface density, color, SFR, and specific SFR \citep[sSFR; \eg][]{catinella10,huang12,gavazzi13}. The positive correlations found between \hi~gas fraction and blue color or sSFR imply a link between galaxies' \hi~content and their current global star formation. 

Starburst galaxies may depart from the typical relations between \hi~content and star formation, however. The Kennicutt-Schmidt Law may differ for starburst galaxies, with starburst galaxies forming stars more efficiently from a given molecular gas mass \citep[\eg][]{kennicutt98,daddi10,genzel10}. One possible explanation for this increased efficiency is merger activity \citep[\eg][]{young86,sanders86,combes94}, and many starbursts appear to be interacting systems. If starbursts gain \hi~gas via major mergers instead of accretion, they may differ from non-starbursts in both their \hi~gas consumption times and \hi~gas fractions. In addition, due to their young stellar populations, mechanical and radiative feedback will have a stronger effect on the ISM of starbursts. This feedback may decrease the \hi~gas fractions of starbursts by driving outflows or ionizing the neutral gas. \citet{oey07}~suggest the latter scenario as an explanation for the lower \hi~gas fractions in starburst galaxies in the Survey for Ionization in Neutral Gas Galaxies (SINGG). The SINGG result contrasts with the ALFALFA and GASS trends of higher \hi~gas fractions in more highly star forming galaxies and demonstrates that the \hi\ content of starbursts requires further study.

Starburst galaxies present an opportunity to study the relations between neutral gas content, star formation, and feedback in extreme conditions. The \hi~gas fractions and kinematics of starbursts may reveal the mechanisms for triggering extreme star formation episodes and the impact of feedback on global gas content. Previous studies of \hi~in starbursts have focused on individual galaxies or optically selected samples \citep[\eg][]{yun93,huchtmeier07,oey07,lopezsanchez12}. To systematically compare the \hi~properties and star formation efficiencies of gas-rich starbursts and gas-rich non-starbursts, we use the ``Fall-sky" portion of the ALFALFA~H$\alpha$ survey. ALFALFA~H$\alpha$~is a volume-limited subset of the ALFALFA survey consisting of 1555 galaxies with follow-up H$\alpha$~and R-band imaging \citep{vansistine15}. With the 565 galaxies in the completed ``Fall sample" of the ALFALFA~H$\alpha$~dataset, we investigate the regulation of the \hi~gas supply throughout the star-formation process.

\section{Data and Methods}
\label{sec:data}

\subsection{The ALFALFA~H$\alpha$~Survey}
\label{sec_survey}
The recently-completed ALFALFA survey is a blind 21-cm survey with Arecibo that covers 7000 deg$^2$~of sky. The \hi-selected ALFALFA~H$\alpha$~survey consists of all ALFALFA-detected galaxies within two designated areas, a Fall-sky region and a Spring-sky region \citep{vansistine15}. The Fall sample only includes galaxies with reliable \hi~detections (\ie ALFALFA codes 1 and 2) and with recession velocities $v=1460-7600$ \kmps. These velocities correspond to distances of $\sim$20-100 Mpc, and the sample is volume-limited for $M_{\rm HI}>10^{9.3}$\Msol. \hi~masses, 21-cm velocity widths, and distances come from the ALFALFA catalog \citep{haynes11}. The selection of the most probable optical counterparts is described in \citet{haynes11}; ambiguous optical identifications or blended \hi~signals from multiple galaxies occur approximately 10\% of the time.

In this work, we consider the complete, ALFALFA~H$\alpha$~Fall sample, which contains 565 galaxies; the full sample is described in \citet{vansistine15}. To compare the global star formation of these galaxies with their global \hi\ content, we obtained $R$-band and H$\alpha$~imaging for the Fall-sky ALFALFA~H$\alpha$~galaxies with the WIYN 0.9m telescope\footnote{The WIYN 0.9m telescope is operated by WIYN Inc.\ on behalf of a Consortium of partner Universities and Organizations (see www.noao.edu/0.9m for a list of the current partners). WIYN is a joint partnership of the University of Wisconsin at Madison, Indiana University, the University of Missouri, and the National Optical Astronomical Observatory.}  and the Kitt Peak National Observatory (KPNO) 2.1m telescope between Sept. 2006 and Oct. 2012. The parameters of the ALFALFA~H$\alpha$~survey, observations, and data reduction are described in \citet{vansistine15}. At the 20-100 Mpc distances of our sample, the observed H$\alpha$ emission typically extends over several tens of arcsec in each galaxy, and the 3\arcsec\ spectroscopic fiber of the Sloan Digital Sky Survey (SDSS) does not accurately capture the total star formation.

H$\alpha$~and $R$-band fluxes were measured for each galaxy individually using aperture photometry. The H$\alpha$~images were continuum-subtracted prior to flux measurement, and $R$-band fluxes were corrected for H$\alpha$\ contamination within the bandpass. Of the 565 Fall sample galaxies, 542 galaxies are detected in H$\alpha$. To facilitate comparison with the literature, the quoted H$\alpha$~equivalent widths are an observed quantity and are not corrected for extinction or \nii~emission. For derived parameters, such as SFRs, the H$\alpha$~fluxes were first corrected for Galactic absorption using the \citet{schlafly11}~recalibration of the \citet{schlegel98}~extinction maps. H$\alpha$~fluxes were then corrected for \nii~contamination and internal absorption using the observed $R$-band absolute magnitudes ($M_R$) and scaling relations derived from a sample of 803 star-forming galaxies from the KPNO International Spectroscopic Survey (KISS; \citealt{salzer00, salzer05}). These corrections are described further in \citet{vansistine15}. We compare these extinction corrections to corrections derived from {\it Wide-field Infrared Survey Explorer }({\it WISE}\footnote{This publication makes use of data products from the {\it Wide-field Infrared Survey Explorer}, which is a joint project of the University of California, Los Angeles, and the Jet Propulsion Laboratory/California Institute of Technology, funded by the National Aeronautics and Space Administration.}; \citealt{wright10})~photometry in \S~\ref{sec_wise}. After correcting the H$\alpha$~emission for Galactic and internal extinction and \nii~emission, we convert the H$\alpha$~luminosities to SFRs. The ALFALFA~H$\alpha$\ SFRs presented in \citet{vansistine15} use the \citet{kennicutt98rev}~calibration, which assumes a \citet{salpeter55} initial mass function (IMF).  Here, we scale the \citet{kennicutt98rev} calibration to a \citet{chabrier03} IMF and calculate the SFR as:
\begin{equation}
{\rm SFR}=4.6 \times 10^{-42} L(\rm{H}\alpha),
\end{equation}
where $L$(H$\alpha$)~is the H$\alpha$~luminosity in \ergps~ and the SFR has units of \Msol~yr$^{-1}$. The \citet{chabrier03} IMF results in SFRs that are a factor of 1.7 lower than SFRs calculated with a \citet{salpeter55} IMF.

\subsection{Stellar Mass Estimation}
\label{sec_masses}
To estimate stellar masses for our sample, we need to account for galaxy-to-galaxy variations in the stellar mass-to-light ratio ($M/L$). \citet{bell01}~demonstrate that galaxy $M/L$~ratios should vary systematically with star formation history and enrichment history, leading to a dependence of $M/L$~on galaxy color. Accounting for the correlation between $M/L$~ratio and color is particularly important to accurately estimate the masses of starburst galaxies, whose young stellar populations amplify the luminosities of all UV-NIR bands. 

\begin{figure*}
\epsscale{0.7}
\plotone{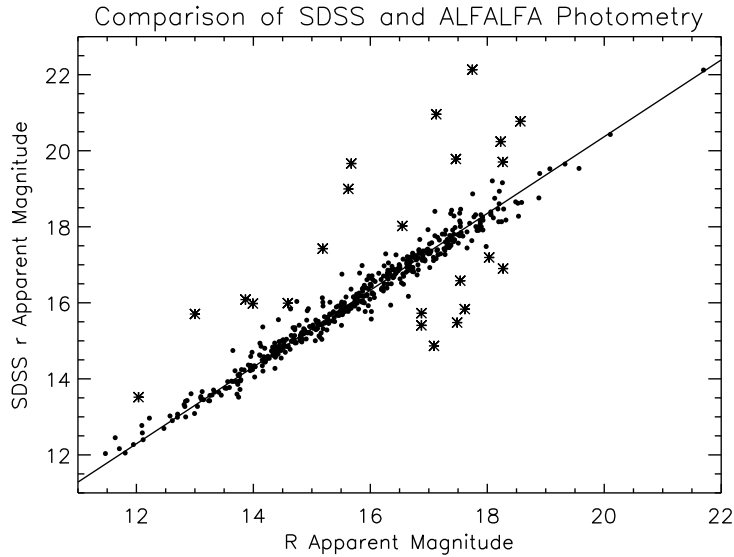}
\caption[A comparison of the SDSS $r$-band and the ALFALFA~H$\alpha$\ $R$-band photometry for galaxies in the Fall sample.]{A comparison of the SDSS $r$-band and the ALFALFA~H$\alpha$\ $R$-band photometry for galaxies in the Fall sample. The line represents a least-squares fit to all data points. Statistical errors are typically smaller than the symbol size. Asterisks indicate galaxies with $r$-band magnitudes that differ from the fitted line by more than 1 mag.}
\label{fig_magcompare}
\end{figure*}

To correct for this effect, we obtain galaxy colors from the SDSS\footnote{Funding for SDSS-III has been provided by the Alfred P. Sloan Foundation, the Participating Institutions, the National Science Foundation, and the U.S. Department of Energy Office of Science. SDSS-III is managed by the Astrophysical Research Consortium for the Participating Institutions of the SDSS-III Collaboration.} Ninth Data Release \citep{ahn12}. SDSS data are available for 513 of the 565 galaxies in the Fall sample. However, SDSS photometry can be problematic, particularly for low surface brightness or irregular galaxies. The SDSS de-blending pipeline separates overlapping objects and sometimes incorrectly shreds one ``parent" galaxy into multiple ``children" \citep[\eg][]{abazajian04, west10}. To determine whether shredding is a concern for our sample, we examine the SDSS data by eye for 10\%~of the ALFALFA~H$\alpha$~galaxies. For 90\% of these galaxies, the $g$ and $r$ photometry of the brightest de-blended child and the parent object agree to within 0.2 mag. We therefore select the brightest de-blended child for each ALFALFA~H$\alpha$~galaxy from the SDSS catalog. As an additional check, for the full ALFALFA~H$\alpha$~Fall sample, we compare the SDSS $r$-band magnitudes with our $R$-band photometry. The photometric data from SDSS and the ALFALFA~H$\alpha$\ $R$-band data show a tight, linear relationship (Figure~\ref{fig_magcompare}), although noticeable outliers exist. An examination of the outliers shows that they are caused by SDSS de-blending errors, overlapping or nearby bright stars, and incorrectly separated galaxy pairs. We treat the SDSS photometry as unreliable if it differs from a least-squares fit to the $R$-band data by more than 1 mag (Figure~\ref{fig_magcompare}). We also eliminate the SDSS photometry for one additional galaxy with a discrepant $g$-band magnitude that results in an unrealistically red $g-r$ color. We do not include these galaxies in any analyses that rely on SDSS-derived parameters, such as stellar masses or galaxy radii. These cuts lead to a sample of 489 ALFALFA~H$\alpha$~galaxies with SDSS photometry. Including the galaxies with uncertain SDSS photometry does not affect any of our conclusions in the following sections.

\begin{figure*}
\epsscale{0.7}
\plotone{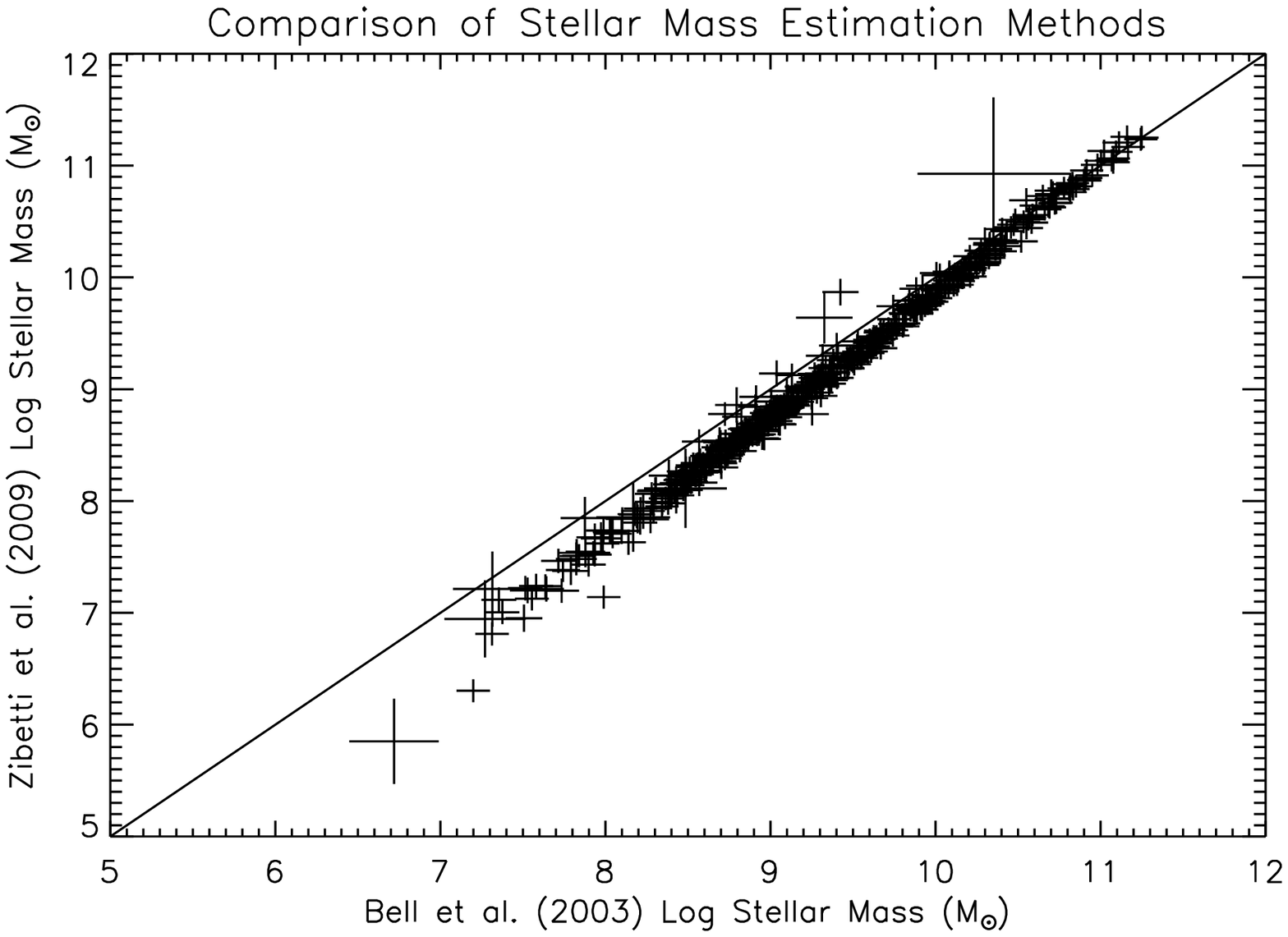}
\caption[A comparison of the stellar masses estimated following \citet{bell03} and \citet{zibetti09}.]{A comparison of the stellar masses estimated following \citet{bell03}~and \citet{zibetti09}. The solid line indicates a one-to-one relation. All masses were estimated using $r$-band luminosities and $g-r$ colors. The errors shown include the statistical photometric errors and the 0.1 dex scatter in the $M/L$-color relation \citep{bell03}. The \citet{zibetti09}~stellar masses are lower, particularly at the low-mass end, due to their adoption of bursty star formation histories.}
\label{fig_zibettibell}
\end{figure*}

We calculate stellar masses using the SDSS $r$-band luminosities and the relationship between $M/L$~ratio and $g-r$ color from Table 7 of \citet{bell03}. We first subtract the observed H$\alpha$~fluxes from the $r$-band fluxes to ensure that nebular emission does not affect the galaxy colors or magnitudes. In addition, we subtract 0.093 dex from the \citet{bell03}~$M/L$~ratios to convert from a ``diet" \citet{salpeter55} IMF \citep[see][]{bell01} to a \citet{chabrier03} initial mass function \citep{gallazzi08,zibetti09}. Systematic uncertainties from the assumed dust extinction and star formation histories dominate the uncertainties in the \citet{bell03} $M/L$~ratios. Following \citet{bell03}, we adopt a total systematic uncertainty of 0.1 dex for the $M/L$~estimates. We also compare the resulting stellar mass estimates with the alternative prescription of \citet{zibetti09}. While \citet{bell03}~assume a smooth star formation history to model galaxy colors, \citet{zibetti09}~consider the effect of bursts. As a result, masses obtained following \citet{zibetti09}~are generally lower than the \citet{bell03}~mass estimates (Figure~\ref{fig_zibettibell}). In particular, at the low-mass end, the stellar masses differ by about a factor of three. However, the relative stellar masses of the galaxies in the sample are only weakly affected. We use the \citet{bell03}~stellar mass estimates for the rest of our analysis, but we note that using the \citet{zibetti09}~mass estimates does not change our conclusions.

\subsection{Selection of Starbursts}
The existing literature contains a variety of definitions as to what constitutes a starburst galaxy \citep[\eg][]{knapen09,bergvall15}. Common starburst definitions are based on high sSFRs or \ha\ equivalent widths (EWs), high sSFRs relative to similar mass galaxies, high SFRs per unit area, or short gas consumption times. In addition, some definitions only consider global star formation enhancements, while others consider smaller-scale enhancements as well. A nuclear starburst, for instance, will have a high SFR in the nuclear region but may not necessarily qualify as a starburst based on the galaxy's total SFR. 

\begin{figure*}
\epsscale{0.75}
\plotone{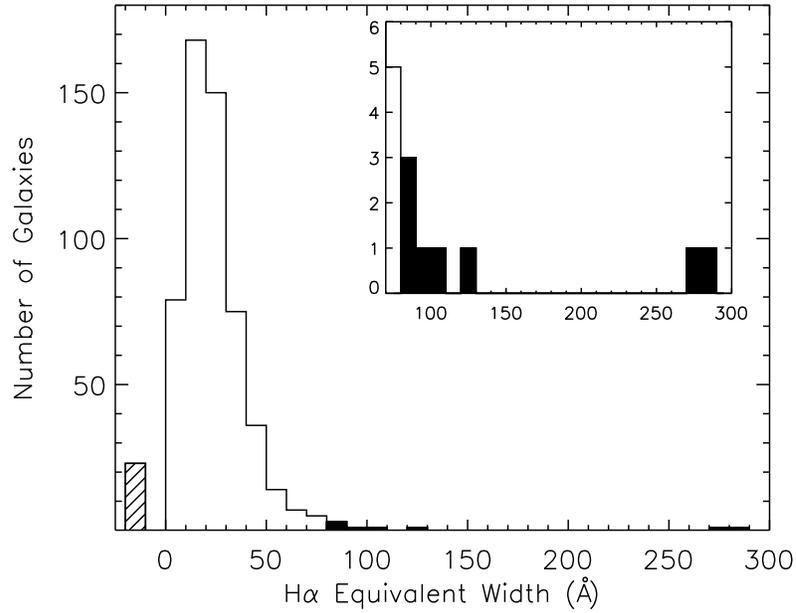}
\caption[The distribution of the H$\alpha$~EWs in the ALFALFA~H$\alpha$~Fall sample.]{The distribution of the H$\alpha$~EWs in the ALFALFA~H$\alpha$~Fall sample. The solid region of the histogram indicates the identified starburst galaxies. The hatched bin at negative EWs represents galaxies without H$\alpha$~detections. The inset shows a zoomed-in view of the high-EW section of the histogram.}
\label{fig_ewhahist}
\end{figure*}

Low-redshift studies often define starbursts as galaxies with high SFRs relative to their past average rates \citep[\eg][]{ostlin01,brinchmann04,lee09,bergvall15}. Here, we adopt a similar definition and define starbursts as galaxies with H$\alpha$~equivalent widths (EWs) greater than 80 \AA. This EW corresponds to a birth rate parameter of $\sim$2, \ie\ the starbursts have instantaneous SFRs greater than or equal to twice their past average SFR \citep{lee09}. Since dust in starburst galaxies may attenuate the ionizing continuum radiation more than the optical stellar continuum, dust effects may lower the observed H$\alpha$~EWs in starbursts \citep[\eg][]{calzetti97,charlot00}. The H$\alpha$~EWs of the starbursts may therefore underestimate their true ratios of current to past star formation. Eight galaxies (1.4\% of the sample) have EWs above the 80 \AA~cut (Figure~\ref{fig_ewhahist}), which is equivalent to an sSFR cut of approximately $6 \times 10^{-10}$ yr$^{-1}$. We list the EWs, SFRs, and \hi~and stellar masses of the starbursts in Table~\ref{sbtable}. 

\begin{table*}
\vspace*{-0.2in}
\begin{center}
\caption{Properties of the ALFALFA~H$\alpha$~Starbursts}
\label{sbtable}
{\scriptsize
\begin{tabular}{lcccccccccc}
\hline
ID & RA$^{a,b}$ & Dec$^{a,b}$ & Distance$^b$ & H$\alpha$~EW & SFR & Log (\mhi)$^b$ & Log ($M_*$)$^c$ & $t_{\rm dep}$ & $A_R^d$ & $W_{20}/W_{50}^b$ \\ 
& & & (Mpc) & (\AA) & (\Msol yr$^{-1}$) & (\Msol) & (\Msol) & (Gyr) &  &  \\
\hline
High EW & & & & & & & & & & \\
Starbursts & & & & & & & & & & \\
\hline
AGC 112546 & 01:31:20.6 & +28:48:29 & 66.4 & 286.6$\pm$4.8 & 0.49$\pm$0.23  & 8.7$\pm$0.1 & 7.2$\pm$0.1 & 1.1$\pm$0.6 & 0.24 & $\leq$1.32 \\
AGC 330517 & 23:32:04.7 & +28:57:21 & 78.4 & 276.4$\pm$4.8 & 6.2$\pm$3.0 & 9.6$\pm$0.1 & 8.0$\pm$0.1 & 0.7$\pm$0.4 & 0.68 & 1.43$\pm$0.43\\
AGC 122866 & 02:11:31.4 & +24:12:51 & 37.3 & 124.5$\pm$4.3 & 0.02$\pm$0.01  & 8.3$\pm$0.1 & 7.6$\pm$0.1 & 8.4$\pm$4.4 & ... & $\leq$1.54 \\
AGC 120193 & 02:22:55.0 & +25:18:53 & 62.7 & 101.9$\pm$2.0 & 0.63$\pm$0.30 & 9.6$\pm$0.1 & ... & 7.1$\pm$3.7 & 0.18 & 1.24$\pm$0.16\\
AGC 333529 & 23:22:39.8 & +28:57:18  & 85.2 & 92.0$\pm$10.0 & 0.08$\pm$0.04 & 9.0$\pm$0.1 & 7.8$\pm$0.1 & 12.8$\pm$6.8 & ... & 1.69$\pm$0.28 \\
AGC 330500 & 23:30:10.0 & +25:32:01 & 82.2 & 81.6$\pm$1.3 & 3.9$\pm$2.0 & 9.9$\pm$0.1 & 9.3$\pm$0.1 & 2.1$\pm$1.1& 0.43 & 1.25$\pm$0.15\\
AGC 122187 & 02:09:55.7 & +27:32:25 & 67.6 & 81.3$\pm$6.1 & 0.05$\pm$0.02 & 9.0$\pm$0.1 & 7.9$\pm$0.1 & 19.3$\pm$10.1 & ... & 1.39$\pm$0.15 \\
AGC 331191 & 23:28:48.9 & +24:52:10 & 71.2 & 80.4$\pm$2.3 & 0.65$\pm$0.31 & 9.6$\pm$0.1 & 9.0$\pm$0.1 & 5.9$\pm$3.1 & 0.18 & 1.62$\pm$0.33\\
\hline
High sSFR & & & & & & & & & & \\
Starbursts & & & & & & & & & & \\
\hline
AGC 122420 & 02:38:50.5 & +27:21:59 & 19.5 & 77.2$\pm$2.1 & 0.03$\pm$0.02 & 8.0$\pm$0.1 & 7.3$\pm$0.1 & 3.2$\pm$1.8 & ... & $\leq$1.33\\
AGC 320466 & 22:57:20.7 & +27:58:52 & 43.3 & 72.7$\pm$8.1 & 0.06$\pm$0.03 & 9.1$\pm$0.1 & 7.7$\pm$0.1 & 23.2$\pm$12.2 & ... & $\leq$1.19\\
AGC 102643 & 00:21:36.7 & +25:28:58 & 95.8 & 66.2$\pm$1.3 & 0.66$\pm$0.32 & 9.3$\pm$0.1 & 9.1$\pm$0.1 & 2.8$\pm$1.5 & 0.29 & 1.13$\pm$0.15\\
AGC 330186 & 23:17:17.7 & +28:36:03 & 96.3 & 23.3$\pm$0.4 & 1.97$\pm$0.98 & 9.6$\pm$0.1 & 9.7$\pm$0.1 & 2.1$\pm$1.1 & ... & $\leq$1.12\\
UGC 470 & 00:44:14.4 & +26:50:35 & 72.7 & 66.9$\pm$1.3 & 2.58$\pm$1.24 & 10.2$\pm$0.1 & 9.8$\pm$0.1 & 5.9$\pm$3.1 & 0.29 & 1.09$\pm$0.01\\
UGC 12821 & 23:52:23.6 & +28:46:15 & 91.7 & 40.7$\pm$0.7 & 10.09$\pm$4.98 & 9.9$\pm$0.1& 10.5$\pm$0.1 & 0.7$\pm$0.4 & 0.34 & 1.40$\pm$0.10\\

\hline
\end{tabular}
\flushleft{
$^a$Coordinates of the detected \hi~source. \\
$^b$Values from the ALFALFA catalog \citep{haynes11}. \\
$^c$The errors listed include the statistical photometric errors and the 0.1 dex scatter in the $M/L$-color relation. \\
$^d$$R$-band asymmetry. \\
}
\break
}
\end{center}
\end{table*}

\begin{figure*}
\epsscale{1.0}
\plotone{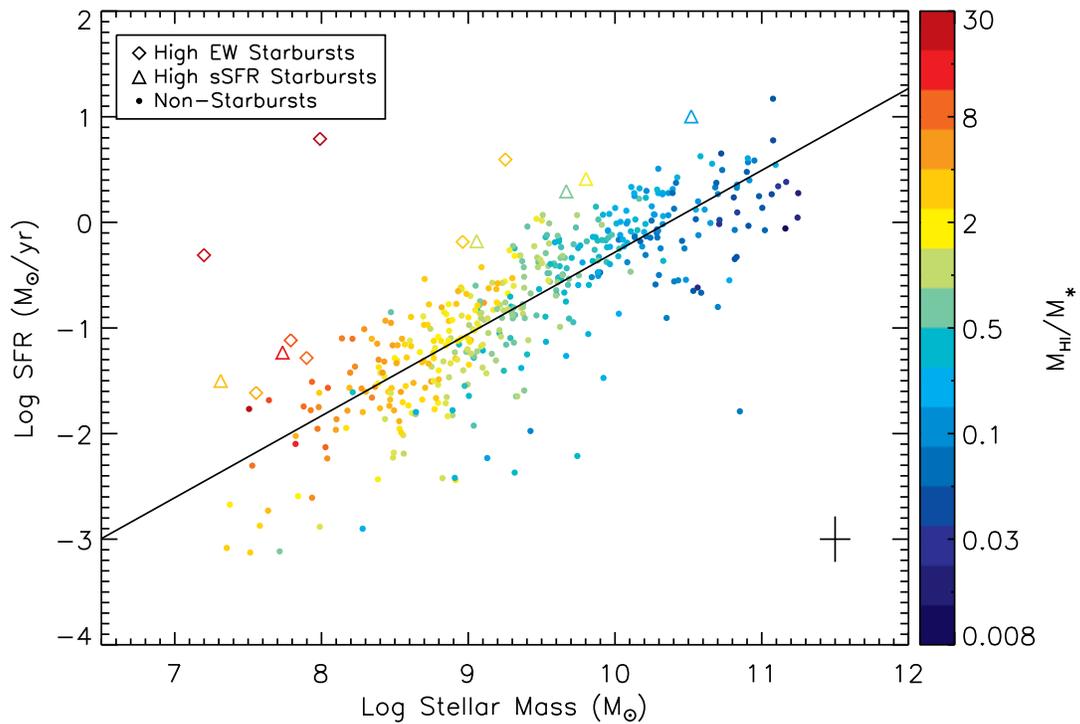}
\caption[SFR, $M_*$, and M$_{\rm HI}$/$M_*$ for the ALFALFA~H$\alpha$~Fall sample.]{SFR and $M_*$~for the ALFALFA~H$\alpha$~Fall sample. Color shows M$_{\rm HI}$/$M_*$, and diamonds indicate starburst galaxies, as identified by H$\alpha$~EW. The black line shows a least-squares fit to the data. Triangles indicate additional starburst candidates, whose SFRs are at least 2$\sigma$\ above the best-fit line. The black cross at the lower right shows representative error bars. At a given $M_*$, galaxies with higher SFRs tend to have slightly higher \hi~gas fractions.  }
\label{fig_gasfrac_mstar_sfr}
\end{figure*}

We also consider an alternative definition of starbursts as galaxies that fall above the main sequence. Figure~\ref{fig_gasfrac_mstar_sfr} shows a least-squares fit line to the SFRs and stellar masses of the sample. An additional six galaxies have SFRs more than 2$\sigma$ above the best-fit line, although none are strong outliers above the main sequence. Table~\ref{sbtable} summarizes the properties of these galaxies. Hereafter, we refer to these six galaxies as the ``High sSFR Starbursts" to distinguish them from the eight ``High EW Starbursts" identified based on H$\alpha$\ EW.

Since we are incomplete below \mhi$=10^{9.3}$\Msol, we may not detect all the starbursts in the Fall-sky volume. For a rough estimate of our completeness, we examine the larger, optically-selected sample of \citet{bothwell09}, which contains 1110 galaxies at distances less than $\sim$43 Mpc. While the \citet{bothwell09} sample selection is less uniform than the ALFALFA~H$\alpha$~sample, it does contain a higher fraction of low-mass galaxies. Assuming the lowest ratios of \mhi/SFR observed for late-type galaxies in the \citet{bothwell09}~sample and converting to a 
\citet{chabrier03} IMF, we should detect all starbursts with $M_*=10^8$ \Msol\ and sSFR$\gtrsim10^{-7}$ yr$^{-1}$ and all starbursts with $M_*=10^9$ \Msol\ and sSFR$\gtrsim10^{-8}$ yr$^{-1}$. However, we may miss any low-mass starburst galaxies whose gas is predominantly molecular. Although the few low-mass starbursts with CO measurements generally appear to have larger \hi~masses than \htwo~masses \citep[\eg][]{kobulnicky95,stil02,bravo04,israel05,nidever13}, we caution that we may not detect the most extreme, low-mass starbursts with the highest \htwo/\hi\ ratios.

\subsection{WISE Data}
\label{sec_wise}
Dust extinction may substantially affect the observed star formation properties of the ALFALFA~H$\alpha$~galaxies. Using infrared measurements from {\it WISE}, we verify the accuracy of the ALFALFA~H$\alpha$~extinction corrections. Later, in \S~\ref{sec:hissfr}, we use the {\it WISE} data to examine the relationship between dust and \hi~content.

We obtain 3.4, 4.6, 12, and 22 $\mu$m fluxes for the sample from the {\it WISE} All-Sky Release Source Catalog \citep{wright10,cutri12}. Elliptical aperture photometry from {\it WISE} is available for sources identified in the Two Micron All Sky Survey (2MASS) Extended Source catalog \citep{jarrett00}. We use this elliptical aperture photometry, where it exists, for objects flagged as resolved sources in the {\it WISE} catalog and the profile fit magnitudes for all other sources. We then match each ALFALFA~H$\alpha$ galaxy to its closest {\it WISE} counterpart within 6\arcsec, the resolution at 3.4 $\mu$m. Following \citet{jarrett13}, we apply the necessary color and magnitude corrections to the {\it WISE} photometry. Of the 565 ALFALFA~H$\alpha$~Fall galaxies, 263 galaxies have detections in all four WISE bands (3.4, 4.6, 12, and 22 $\mu$m) and an additional 95 galaxies have detections in three WISE bands (3.4, 4.6, and 12 $\mu$m). 

The extinction corrections adopted for the ALFALFA~H$\alpha$~galaxies in \S~\ref{sec_survey} are based on the galaxies' $R$-band luminosities. Since this is a statistical correction, it may not be correct on an individual galaxy basis. To test the adopted extinction correction, we use the {\it WISE} and H$\alpha$~observations to estimate extinction-corrected SFRs for individual galaxies. \citet{wen14}~ calibrate the WISE 12 $\mu$m~and 22 $\mu$m~bands as extinction indicators using the H$\alpha$/H$\beta$~ratios from SDSS spectra of $z<0.25$~star-forming galaxies. Following \citet{wen14}, we calculate
\begin{equation}
\label{eqn_sfrwise}
{\rm SFR}=0.87 \times 10^{-41.27} L_{\rm H\alpha,obs} + a \nu L_\nu,
\end{equation} where $L_{\rm H\alpha,obs}$~is the H$\alpha$~luminosity without an internal extinction correction, $\nu L_\nu$~is the appropriate {\it WISE} luminosity. The coefficient $a$~depends on the {\it WISE} band and extinction law adopted and ranges from $\sim$0.02-0.04. The factor of 0.87 in the equation converts from the \citet{kennicutt12} SFR calibration, which uses a \citet{kroupa03}~IMF, to the \citet{kennicutt98rev} calibration and the \citet{chabrier03}~IMF used in \S~\ref{sec_survey}. We calculate extinction-corrected SFRs using the {\it WISE} 12 $\mu$m and 22 $\mu$m luminosities and assuming a \citet{calzetti00}~extinction law. 

\begin{figure*}
\epsscale{0.7}
\plotone{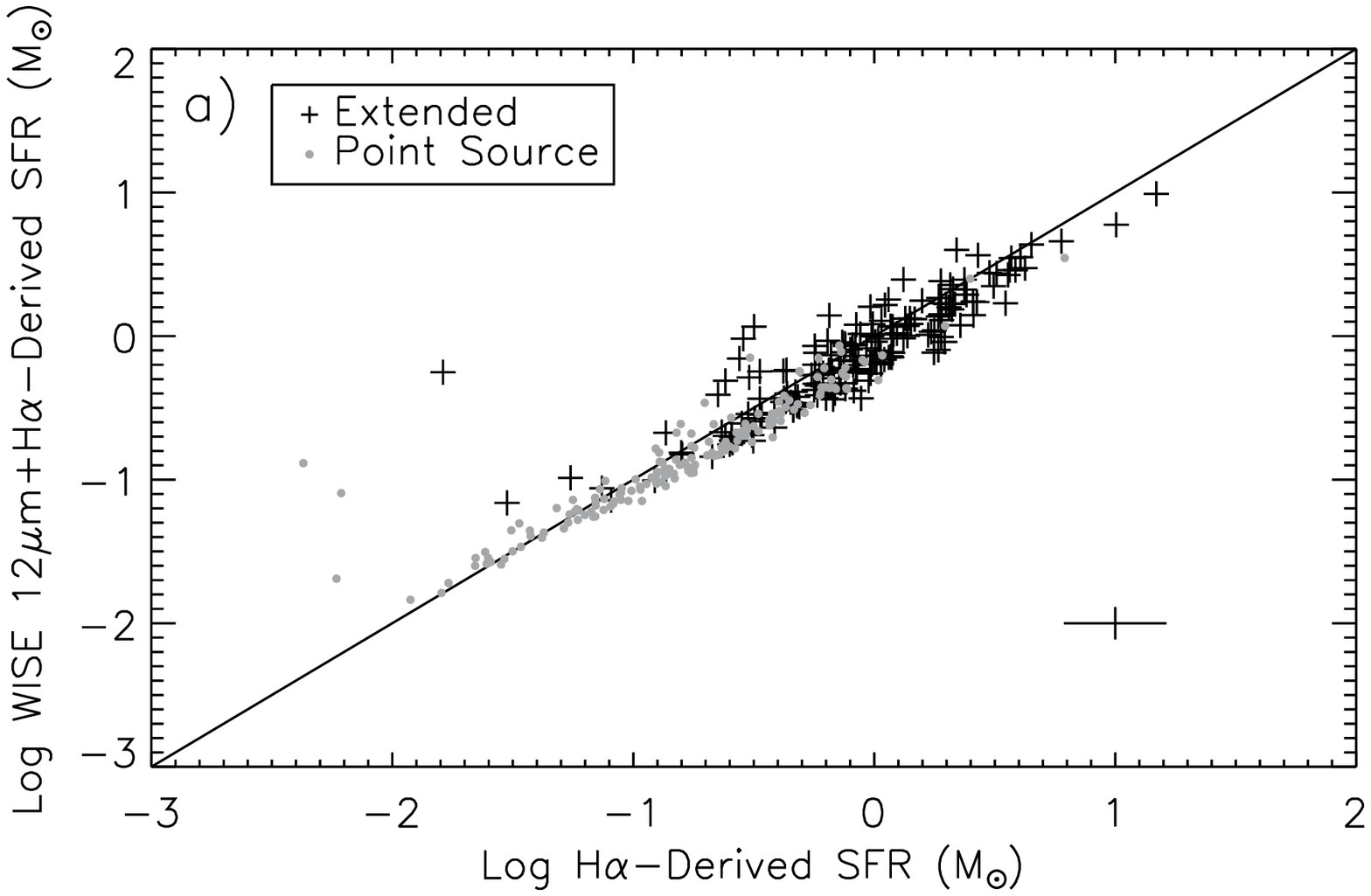}
\epsscale{0.7}
\plotone{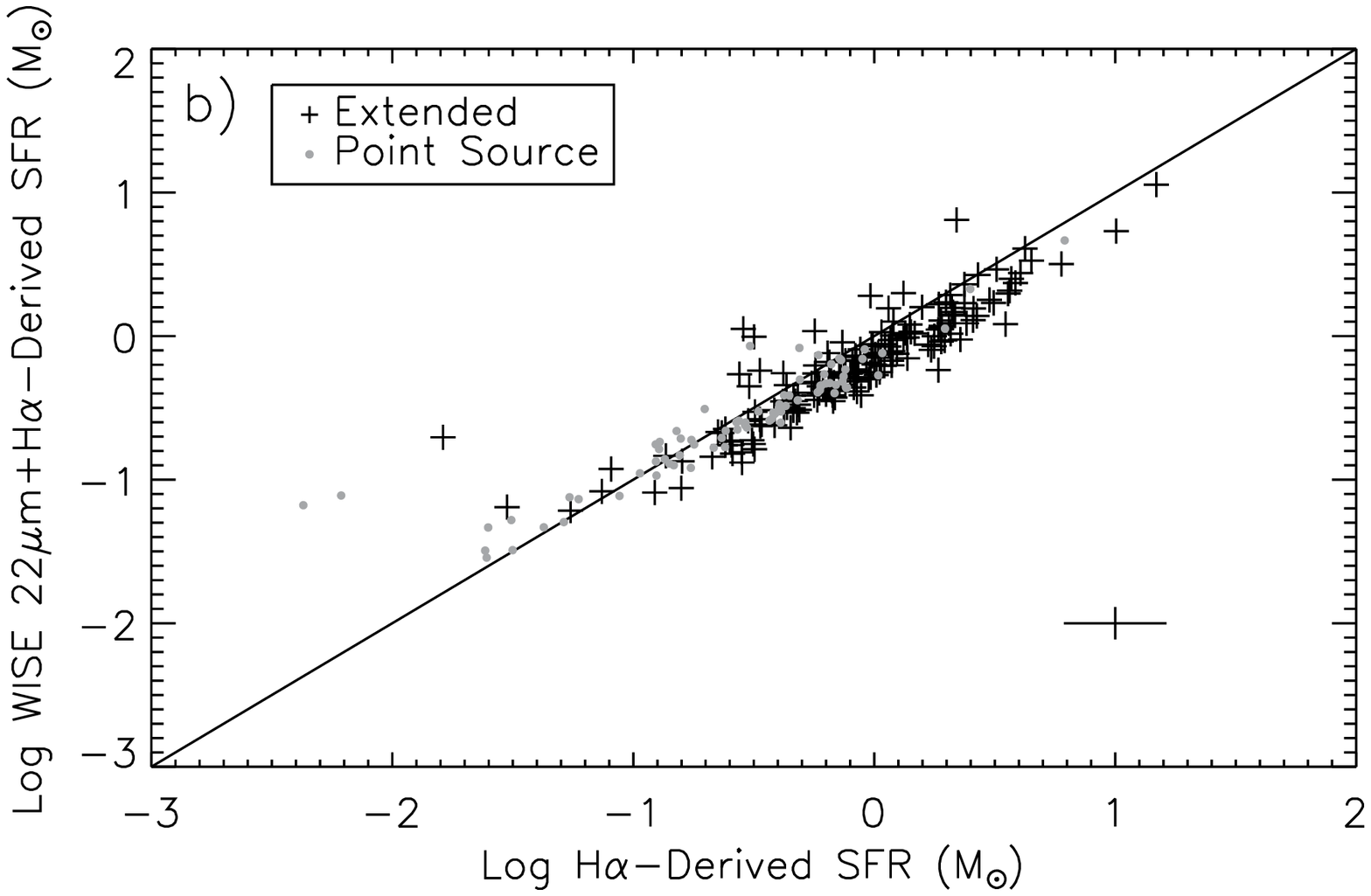}
\caption[A comparison of the ALFALFA~H$\alpha$~SFRs with SFRs corrected for extinction using (\emph {a}) {\it WISE} 12 $\mu$m luminosities and (\emph {b}) 22 $\mu$m luminosities.]{A comparison of the ALFALFA~H$\alpha$~SFRs with SFRs corrected for extinction using (\emph {a}) {\it WISE} 12 $\mu$m luminosities and (\emph {b}) 22 $\mu$m luminosities. Gray circles indicate measurements using {\it WISE} point source photometry, and black crosses indicate galaxies with elliptical aperture photometry. The solid line shows a one-to-one relation. Representative error bars are shown in the lower right corner. }
\label{fig_wisecompare}
\end{figure*}

We compare these SFRs with the ALFALFA~H$\alpha$~SFRs in Figure~\ref{fig_wisecompare}. Overall, the {\it WISE}-derived SFRs and ALFALFA~H$\alpha$~SFRs show reasonable agreement. A few galaxies show higher SFRs using the {\it WISE} extinction correction than the original ALFALFA~H$\alpha$ extinction correction. These galaxies show prominent dust lanes or appear reddened in SDSS images. For these particular objects, the original extinction correction used to derive the ALFALFA~H$\alpha$~SFRs is likely insufficient. The {\it WISE}-derived SFRs appear systematically lower than the ALFALFA~H$\alpha$~SFRs by $\sim$0.1 dex for many galaxies, especially point-source {\it WISE} detections at moderate SFRs. The {\it WISE} profile-fit fluxes for these sources may be underestimated, as discussed in \citet{cutri12b}. In general, however, the {\it WISE}-based extinction corrections agree well with the adopted ALFALFA~H$\alpha$\ extinction corrections.

\section{Results}
\label{sec:results}
\subsection{\hi~Gas Content}
\label{sec:hi}
\subsubsection{The \hi~Gas Supply of Starbursts}
\label{sec:hisbs}

Star formation is intimately linked with the cold gas content of galaxies. A large \hi~supply may be necessary to fuel high SFRs, but the resulting feedback may expel or ionize much of the \hi~gas. Given the extreme levels of star formation and feedback in starburst galaxies, we consider whether their \hi~content differs from the other galaxies in the ALFALFA~H$\alpha$~sample. High \hi~gas fractions in starbursts may suggest that a large gas reservoir is a key precondition for triggering a starburst, while \hi-deficiencies might indicate that radiative feedback plays the dominant role in shaping starbursts' ISM. 

In non-starbursts, larger \hi~reservoirs do appear to lead to enhanced star formation. As shown by \citet{huang12} for ALFALFA galaxies, SFR correlates with \mhi\ over almost 4 dex in \hi\ mass, indicating a link between atomic gas and star formation. Since part of this trend may result from the tight relation between SFR and galaxy stellar mass, we examine whether highly star-forming galaxies have more \hi~than galaxies of a similar mass. In Figure~\ref{fig_gasfrac_mstar_sfr}, we show the ratio of \mhi~to stellar mass ($M_*$) as a function of SFR and $M_*$. For consistency with previous \hi\ studies \citep[\eg][]{catinella10, huang12}, we refer to \mhi/$M_*$~ as the \hi\ gas fraction. Previous studies have found that at a given stellar mass, galaxies with higher SFRs tend to be more \hi-rich \citep[\eg][]{wang11, huang12}, and the ALFALFA~H$\alpha$~sample likewise exhibits this trend. The larger, 40\%\ complete ALFALFA sample \citep{huang12} shows that this relation between SFR and \hi\ content also holds at lower \hi\ gas fractions than probed by the ALFALFA~H$\alpha$\ galaxies. However, while some of the starbursts' SFRs are an order of magnitude higher than other galaxies of the same mass, their \hi~gas fractions do not show a comparable increase. In fact, the starburst \hi~gas fractions are similar to those of star-forming galaxies with much lower SFRs. In the starburst regime, \hi~content and star formation do not appear to be closely coupled.

\begin{figure*}
\epsscale{0.7}
\plotone{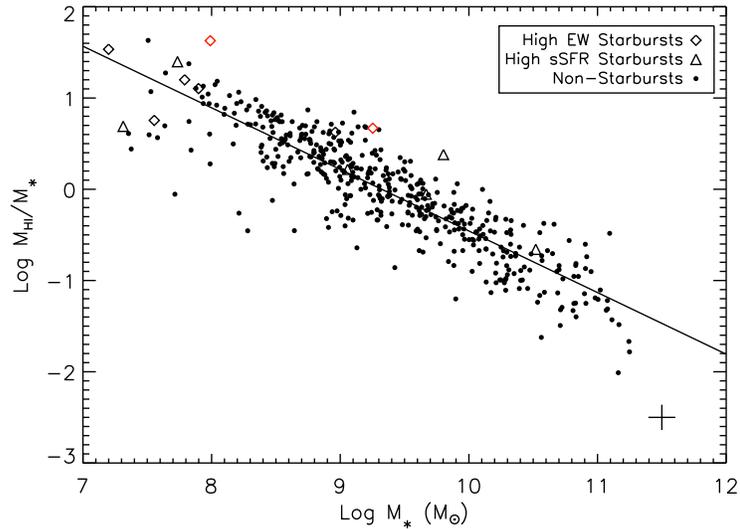}
\caption[$M_{\rm HI}/M_*$ and $M_*$~for the ALFALFA~H$\alpha$~Fall sample.]{$M_{\rm HI}/M_*$ and $M_*$~for the ALFALFA~H$\alpha$~Fall sample. Symbols are described in Figure~\ref{fig_gasfrac_mstar_sfr}. The solid line shows a least-squares fit to the non-starburst galaxies. Two high-EW starbursts, shown in red, have $M_{\rm HI}/M_*$ values 2$\sigma$~higher than the best-fit line. These starbursts are discussed further in \S~\ref{sec_merge}. }
\label{fig_gf_mstar}
\end{figure*}

The starbursts are \hi-rich relative to the ALFALFA~H$\alpha$~sample as a whole, but this \hi-richness results from their lower-than-average stellar masses. Low-mass galaxies in general tend to be more \hi-rich \citep[\eg][]{gavazzi96,huang12}, and in most cases, the \hi~gas fractions of the starbursts are similar to other galaxies of the same mass. The relatively high $M_{\rm HI}$ detection limit of the ALFALFA~H$\alpha$ survey leads to a tight apparent relation between $M_{\rm HI}/M_*$ and $M_*$; Figure~\ref{fig_gf_mstar} displays the least-squares fit to this relation for the non-starbursts in the sample. Most of the starbursts have $M_{\rm HI}/M_*$ values that are within 1$\sigma$ of the best-fit line. Only two of the high EW starbursts and one high sSFR starburst have $M_{\rm HI}/M_*$ ratios that appear high for their stellar mass, deviating by $>2\sigma$~from the main trend (Figure~\ref{fig_gf_mstar}). Two starbursts even have lower-than-average \hi~gas fractions. This fact suggests that the high SFRs of the starbursts are not caused by an excess of \hi, but rather by an enhanced efficiency of converting \hi~into molecular gas and stars.

\begin{figure*}
\epsscale{1.25}
\plotone{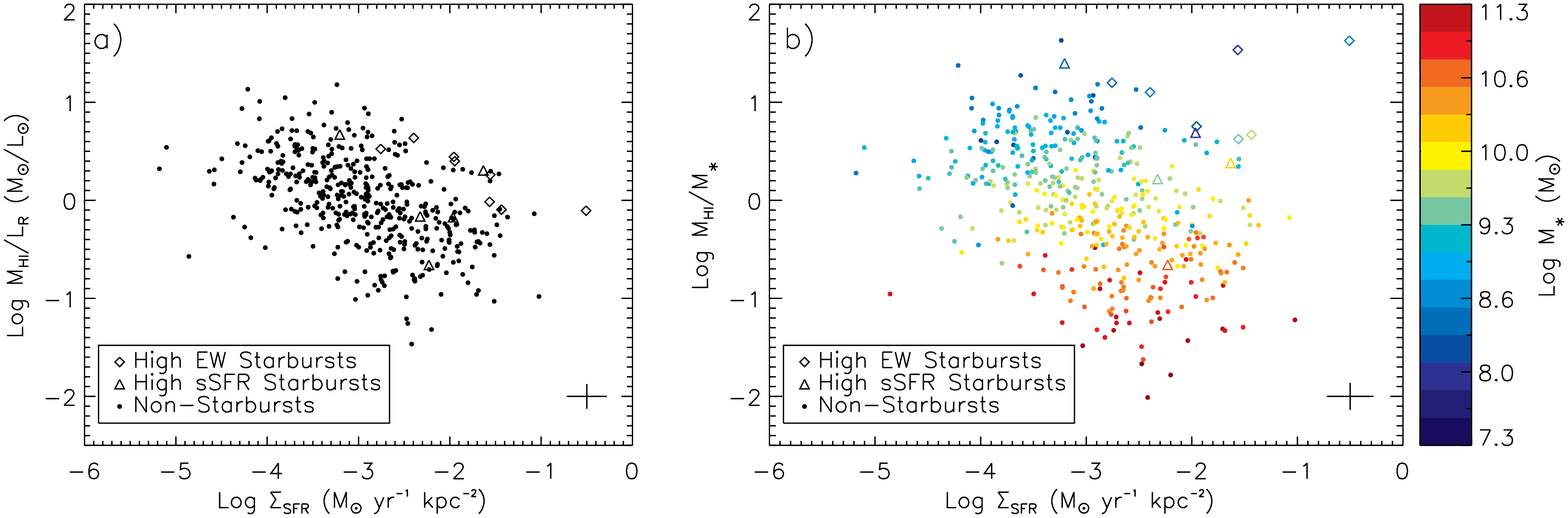}
\caption[(\emph {a}) $M_{\rm HI}$/$L_R$~vs.~$\Sigma_{\rm SFR}$ for comparison with the SINGG sample. (\emph {b}) $M_{\rm HI}$/$M_*$~vs.~ $\Sigma_{\rm SFR}$.]{(\emph{a}) \mhi/$L_R$~vs.~$\Sigma_{\rm SFR}$, with symbols as in Figure~\ref{fig_gasfrac_mstar_sfr}. Galaxies with higher $\Sigma_{\rm SFR}$ have lower \mhi/$L_R$. (\emph{b}) When we substitute $M_*$~estimates from SDSS photometry for $L_R$, \gf~shows a weaker negative correlation with $\Sigma_{\rm SFR}$, and high $\Sigma_{\rm SFR}$ galaxies span a wide range of \gf. Color shows $M_*$.}
\label{fig_singg}
\end{figure*}

Although the starbursts are not excessively \hi-rich for their stellar masses, they are also not \hi-deficient. This conclusion contrasts with the results of \citet{oey07} for starbursts in the SINGG sample. However, \citet{oey07} use $R$-band magnitude as a proxy for stellar mass and identify starbursts based on H$\alpha$ surface brightness, rather than EW. To compare with the SINGG results, we calculate the SFR surface density for a flat disk morphology as 
\begin{equation}
\label{sfi_eqn}
{\Sigma_{\rm SFR}}=\frac{{\rm SFR_{50}}}{2 \pi R_{50}^2},
\end{equation} where $R_{50}$~is the $R$-band half-light radius and SFR$_{50}$~is the SFR within that radius. We show the relation between $\Sigma_{\rm SFR}$ and the ratio of \mhi~to $R$-band luminosity, \mhi/$L_R$, in Figure~\ref{fig_singg}a. Our sample spans the same range of \mhi/$L_R$ as the SINGG sample, and we likewise see that galaxies with higher $\Sigma_{\rm SFR}$ do appear to have lower ratios of \mhi/$L_R$. On the other hand, as discussed in \S~\ref{sec_masses}, $R$-band luminosities systematically overestimate the stellar masses of starbursts, leading to underestimates of their \hi~gas fractions. In Figure~\ref{fig_singg}b we replace $L_R$~with $M_*$, calculated as described in \S~\ref{sec_masses}. The trend between \hi~content and $\Sigma_{\rm SFR}$ now appears weaker, and even at the highest $\Sigma_{\rm SFR}$ end, galaxies span a range of \hi~gas fractions. This result highlights the importance of multicolor photometry for accurate stellar masses and gas fractions. The weak residual negative correlation between \gf~and $\Sigma_{\rm SFR}$ in Figure~\ref{fig_singg}b is due to the scalings of \hi~gas fraction, SFR, and half-light radius with $M_*$~in the ALFALFA~H$\alpha$~mass regime \citep[\eg][]{catinella10, huang12, brinchmann04, shen03}. The large range of gas fractions among high $\Sigma_{\rm SFR}$ galaxies shows that feedback from starbursts does not necessarily result in a reduced \hi~content.

The \hi-richness of the starbursts indicates that they are not able to completely ionize or consume their neutral gas. Radiative feedback in high $\Sigma_{\rm SFR}$ galaxies does not appear to significantly affect their global \hi~gas fractions, perhaps because most of the \hi~gas mass resides at radii well outside the star-forming regions. The gas column near the starburst may be thick enough to prevent the escape of ionizing photons, and in addition, any decrease in \hi~from photoionization may be offset by an increase in \hi~from \htwo~photodissociation.

\subsubsection{The Connection Between \hi~Gas Fraction and Specific Star Formation Rate}
\label{sec:hissfr}

\begin{figure*}
\epsscale{1.0}
\plotone{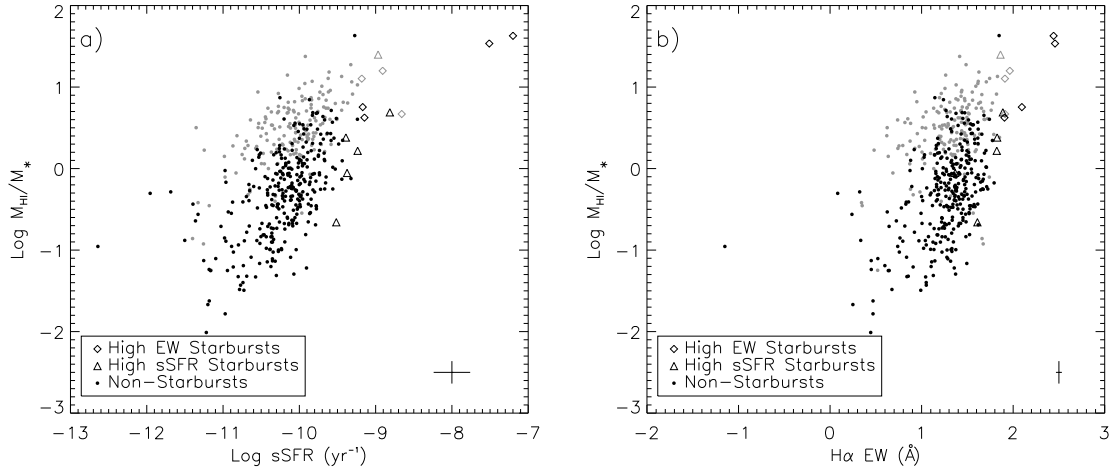}
\caption[(\emph {a})$M_{\rm HI}$/$M_*$~vs.~sSFR. (\emph {b}) $M_{\rm HI}$/$M_*$~vs.~H$\alpha$~EW.]{(\emph{a})\gf~vs.~sSFR, with symbols as in Figure~\ref{fig_gasfrac_mstar_sfr}. To compare with Figure~\ref{fig_gf_color}b, we show ALFALFA~H$\alpha$~galaxies detected in {\it WISE} in black and non-detections in gray. (\emph{b}) \gf~vs.~H$\alpha$~EW.}
\label{fig_gf_ssfr}
\end{figure*}

\begin{figure*}
\epsscale{0.7}
\plotone{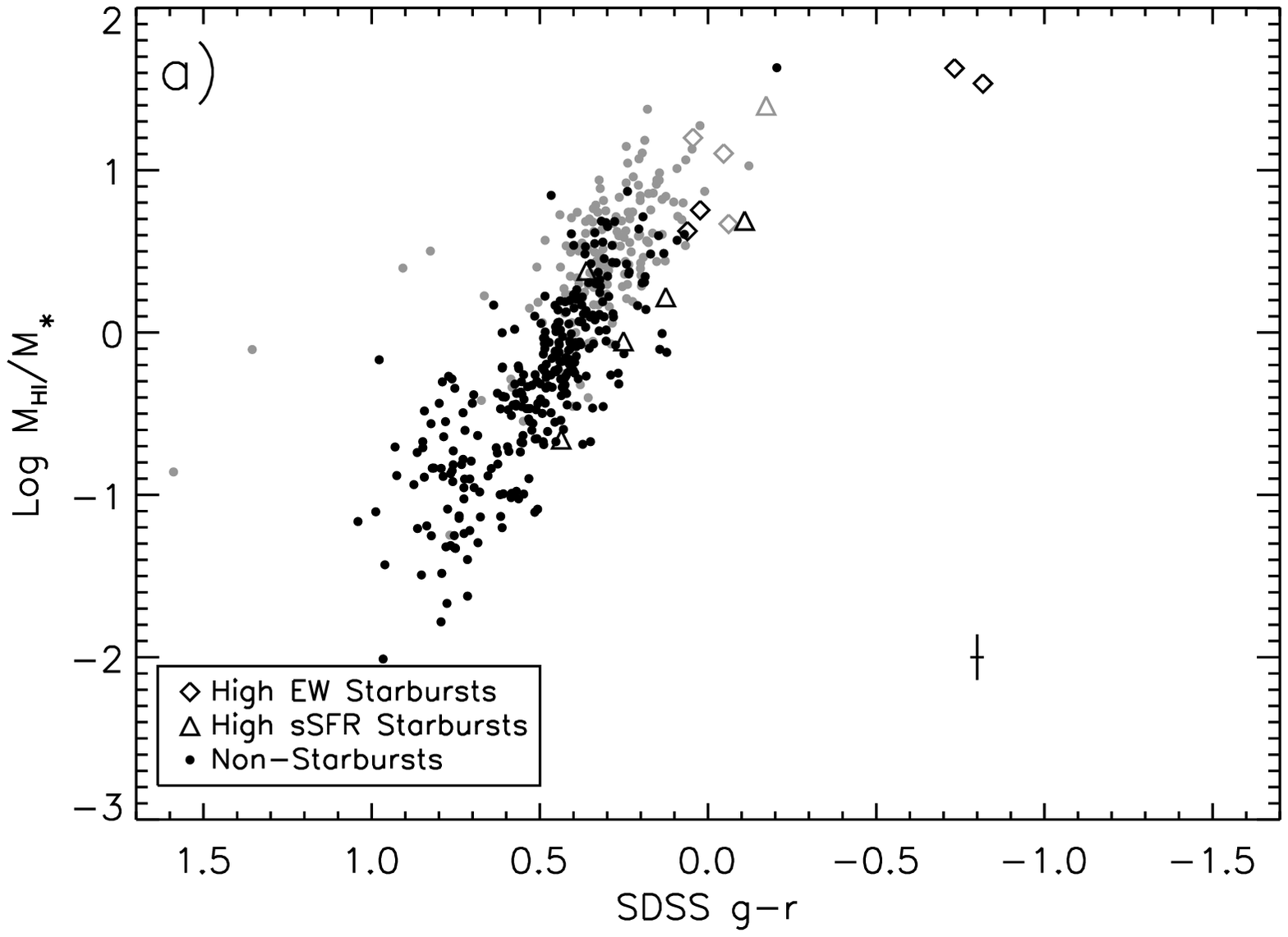}
\epsscale{1.0}
\plotone{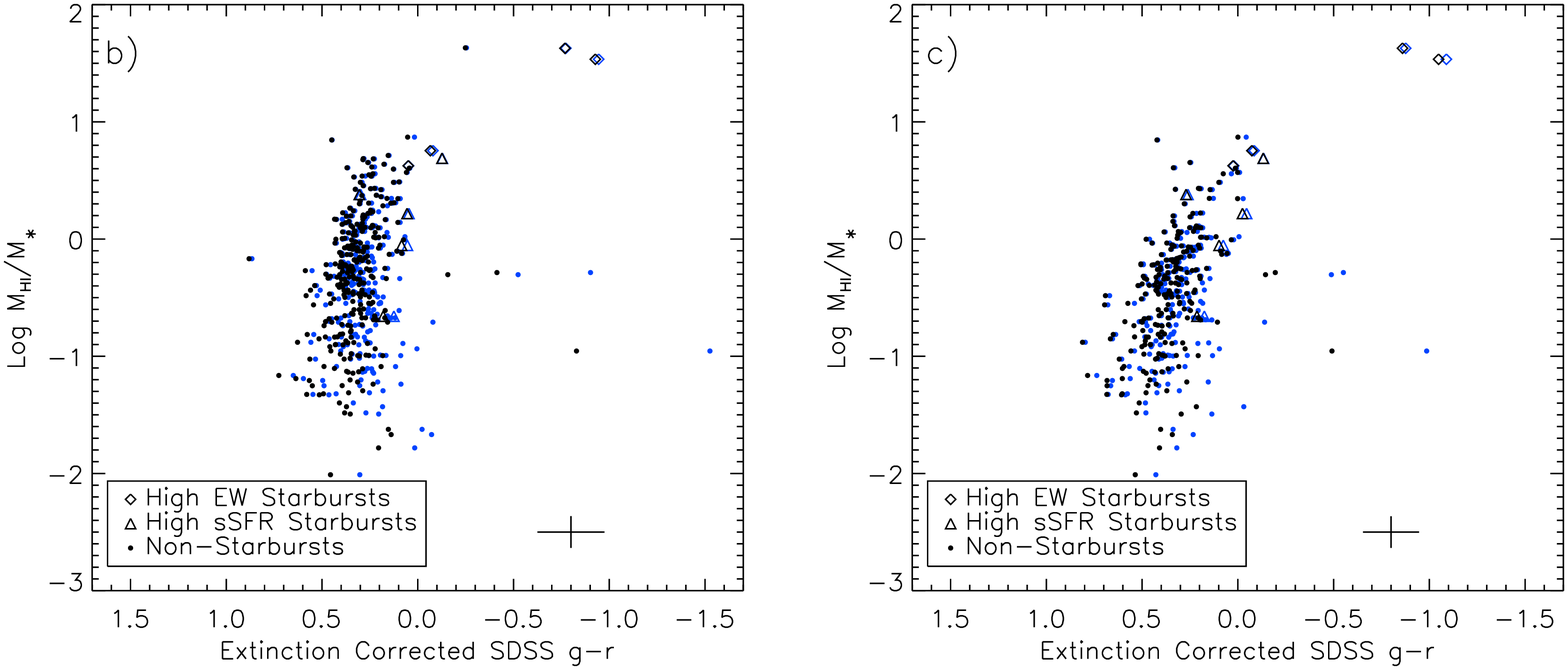}
\caption[(\emph {a}) $M_{\rm HI}$/$M_*$~vs.~SDSS $g-r$~color. (\emph {b}) $M_{\rm HI}$/$M_*$~vs.~SDSS $g-r$~color, after correcting for extinction using the {\it WISE} 12 $\mu$m luminosities. (\emph {c}) $M_{\rm HI}$/$M_*$~vs.~SDSS $g-r$~color, after correcting for extinction using the {\it WISE} 22 $\mu$m luminosities.]{(\emph{a}) SDSS $g-r$~color and \gf~show a tight correlation. Symbols and colors are the same as in Figure~\ref{fig_gf_ssfr}. The $x$-axis is flipped to match the orientation of Figure~\ref{fig_gf_ssfr}. (\emph{b}) \gf~vs.~SDSS $g-r$~color, after correcting for extinction using the {\it WISE} 12 $\mu$m luminosities. Black points indicate extinctions derived assuming a \citet{calzetti00}~extinction law, and blue points indicate extinctions derived with the \citet{cardelli89}~extinction law. The calculated errors include photometric uncertainties, the uncertainties in the coefficients in Equation~\ref{ebv_eqn}~due to adopting different extinction laws, and the uncertainty in the $E(B-V)$~to $E(g-r)$~conversion. (\emph{c}) \gf~vs.~SDSS $g-r$~color, after correcting for extinction using the {\it WISE} 22 $\mu$m luminosities. }
\label{fig_gf_color}
\end{figure*}

The ALFALFA~H$\alpha$~galaxies indicate that \hi~gas fraction generally increases with sSFR, although the data show substantial scatter and this trend may flatten in the starburst regime. The relatively weak connection between sSFR or H$\alpha$~EW and \hi~gas fraction (Figure~\ref{fig_gf_ssfr}) is seemingly at odds with the well-established tight correlation between \hi~gas fraction and NUV-$r$~color \citep[\eg][]{catinella10,huang12}. For instance, the GASS sample of high-mass galaxies shows that the best predictors of a galaxy's \hi~gas fraction are NUV-$r$~color and stellar mass surface density \citep{catinella10}. Similarly, we observe a tight trend between \hi~gas fraction and SDSS $g-r$ color (Figure~\ref{fig_gf_color}a). The weaker trend with H$\alpha$-derived sSFR and strong trend with $g-r$ suggest two possible explanations: (1) dust extinction drives the close correlation between $g-r$~color and \hi~or (2) galaxies' global \hi~content is more closely linked with the SFR averaged over long timescales than with instantaneous star formation \citep[\eg][]{kannappan13}. 

To estimate the effect of dust extinction on SDSS color, we use the {\it WISE} 12 $\mu$m and 22 $\mu$m luminosities (see \S~\ref{sec_wise}) to derive the $E(g-r)$~color excess. Following \citet{wen14}, we calculate $E(B-V)$~using
\begin{equation}
\label{ebv_eqn}
\frac{\nu L_{\nu}}{L_{\rm H\alpha (obs)}}=\frac{10^{0.4k_{\rm H\alpha }\times E(B-V)}-1}{a}
\end{equation} where $\nu L_{\nu}$~is the 12 $\mu$m or 22 $\mu$m luminosity, $L_{\rm H\alpha(obs)}$~is the H$\alpha$~luminosity with no correction for internal extinction, $k_{\rm H\alpha}$~is the reddening curve at H$\alpha$, and $a$~is a coefficient that depends on the extinction law and {\it WISE} band used \citep{wen14}. We then convert $E(B-V)$ to $E(g-r)$, using the conversion in \citet{yuan13}. The median $A_V$~values inferred for the ALFALFA~H$\alpha$ sample using the \citet{calzetti00}~and \citet{cardelli89}~extinction laws range from 0.45-0.6. These values are comparable to the face-on $A_V$~values calculated from radiative transfer models of spiral galaxies \citep{xilouris99,bianchi07,degeyter14}. In Figures~\ref{fig_gf_color}b and \ref{fig_gf_color}c, we show the relation between \mhi/$M_*$~and extinction-corrected $g-r$ color. After correcting the SDSS $g-r$ colors for extinction, the trend with \hi~gas fraction weakens substantially (Figure~\ref{fig_gf_color}). 

Dust extinction has the most dramatic effect on the galaxy color of the red, low sSFR galaxies in our sample. We therefore run \citet{bruzual03}~stellar population synthesis (SPS) models to confirm whether the above dust extinction corrections are realistic for low sSFR galaxies. We use the Padova 1994 isochrones at metallicities of 0.2 and 1 \Zsol~\citep{bressan93, fagotto94} and the \citet{chabrier03}~initial mass function.  To estimate the expected colors of the low sSFR galaxies, we adopt a log-normal star formation history, as recommended by \citet{gladders13}~for field galaxies. For present-day sSFRs of $10^{-12}-10^{-11}$ yr$^{-1}$ and peak star formation at $z=0.7-3$, the intrinsic $g-r$~colors range from 0.5 to 0.8. While not exhaustive, these models hint at the range of colors we should expect for the reddest galaxies in the absence of dust. 

The SPS models cannot reproduce the reddest observed $g-r$~colors, $(g-r) > 0.8$, even assuming peak star formation at $z \ge 6$. In fact, as suggested by the extinction-corrected colors in Figure~\ref{fig_gf_color}, a wide range of star formation histories produces a relatively narrow range of stellar population colors. Colors redder than this range would then result from the effects of dust extinction. The {\it WISE} extinction-corrected colors in Figure~\ref{fig_gf_color}~do appear more blue than predicted by the SPS models, with median $g-r$~colors of 0.2-0.3 using the 12 $\mu$m band and 0.4-0.5 using the $22\mu$m band. Although {\it WISE} is less sensitive at $22\mu$m than at $12\mu$m, the warm dust traced by the 22 $\mu$m band \citep{jarrett13} may be a better indicator of the total stellar extinction than the polycyclic aromatic hydrocarbons (PAHs) traced by the $12\mu$m band. Adopting a more realistic, bursty star formation history in the models could also lead to better agreement with the {\it WISE} extinction-corrected colors. Regardless, both the {\it WISE} data and the SPS models suggest that dust extinction could be responsible for the tight correlation between $g-r$~color and \hi~gas fraction.

\begin{figure*}
\epsscale{0.75}
\plotone{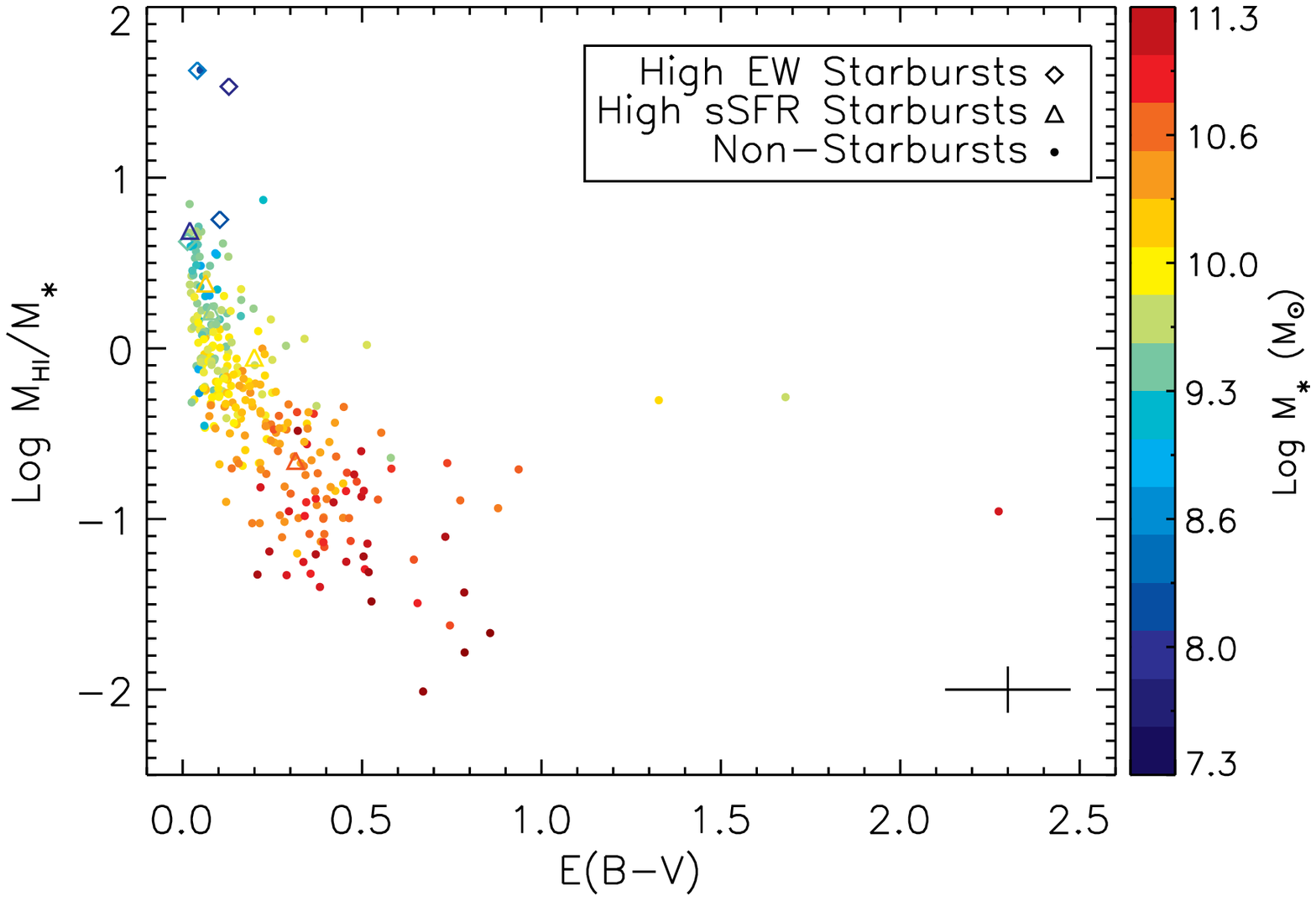}
\caption[$M_{\rm HI}$/$M_*$~vs.~$E(B-V)$.]{\gf~vs.~$E(B-V)$ derived using the {\it WISE} 12 $\mu$m luminosities and \citet{cardelli89}~extinction law. Color indicates stellar mass.}
\label{fig_ebv_gf}
\end{figure*}

Figures~\ref{fig_gf_ssfr} and~\ref{fig_gf_color} show that, in star-forming galaxies, \hi~gas fraction correlates weakly with sSFR; the tight connection between \hi~and $g-r$~color may stem almost entirely from a link between dust extinction and \hi~gas fraction. Nevertheless, we caution that ALFALFA misses most galaxies on the red sequence \citep{huang12}, and including this population may result in a clearer trend between color and \hi~gas fraction \citep[\cf][]{gavazzi13}. 

For our sample, the correlation between \hi~gas fraction and color may reflect an underlying trend between dust content and stellar mass. Figure~\ref{fig_ebv_gf} demonstrates that the {\it WISE}-derived extinctions from the 12 $\mu$m luminosities and \citet{cardelli89}~extinction law correlate with stellar mass and anti-correlate with \hi~gas fraction. The results for the {\it WISE} 22 $\mu$m~data and for the \citet{calzetti00}~extinction law are similar. The outlier at the highest $E(B-V)$~in Figure~\ref{fig_ebv_gf} is UGC 1488, which has a high inclination, prominent dust lane, and signs of morphological disturbance. Higher mass, star-forming galaxies have both higher dust masses and higher dust-to-gas ratios, due to the galaxies' higher average metallicities. Conversely, these same high-mass, dusty galaxies have low \hi~gas fractions, due to higher \htwo\ conversion and star formation efficiencies or lower gas accretion rates relative to their gas consumption rates \citep[\eg][]{dave11b, kannappan13}.

The metallicities of galaxies set their dust content and ultimately drive the observed trend between $g-r$ color and \hi~gas fraction. \citet{bothwell13}~show that the mass-metallicity relation of galaxies depends strongly on \hi~content. As a galaxy evolves, the balance of \hi~inflows, star formation, and outflows determines both its metallicity and its \hi~gas fraction. The tight trend between color and \hi~gas fraction is therefore a manifestation of the key role of \hi\ gas flows in regulating galaxy metallicities.

\begin{figure*}
\epsscale{0.75}
\plotone{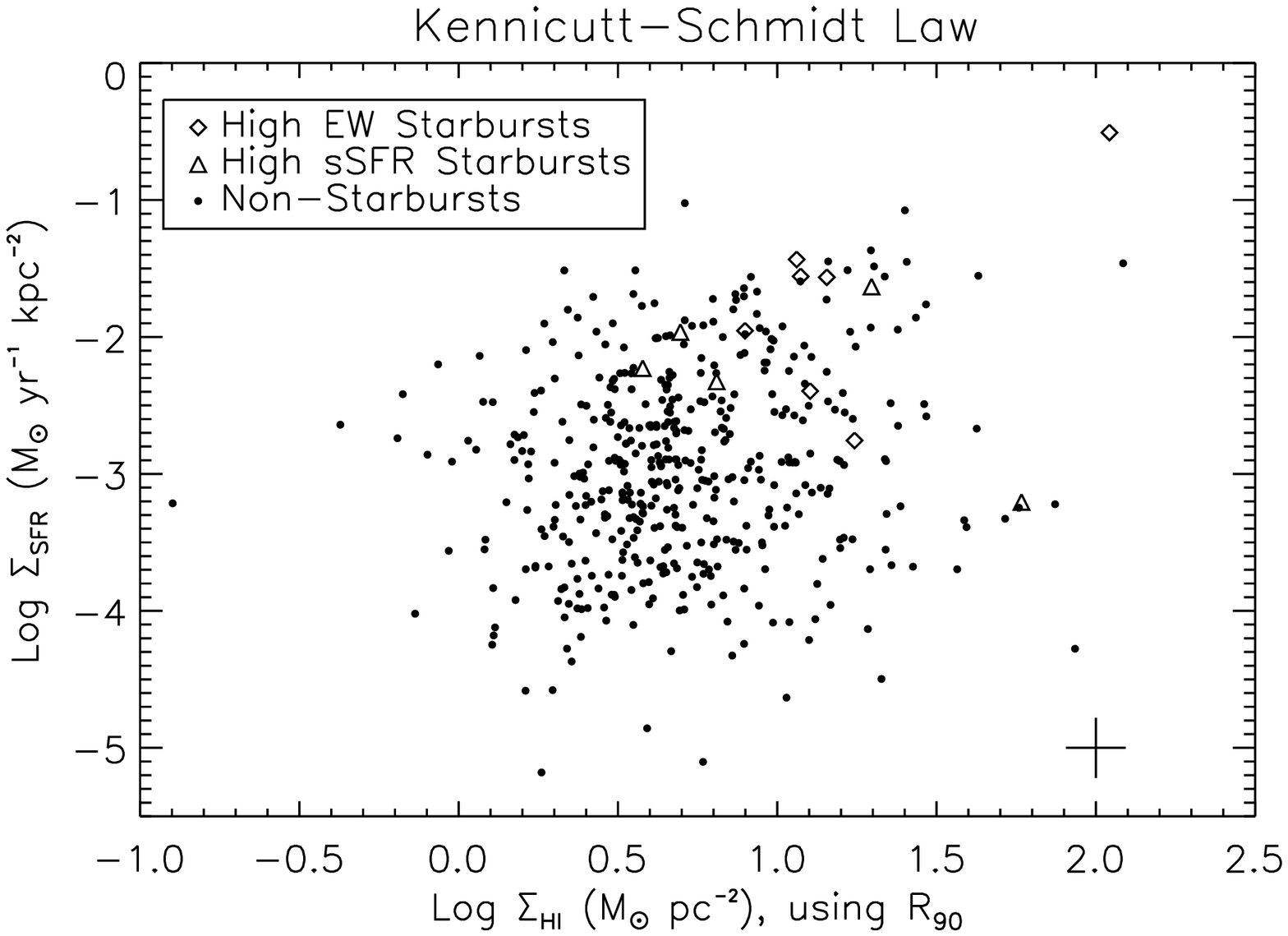}
\caption[The Kennicutt-Schmidt Law for the ALFALFA~H$\alpha$~sample.]{$\Sigma_{\rm SFR}$~vs.~$\Sigma_{\rm HI}$, calculated from Equations~\ref{sfi_eqn}~and \ref{hisd_eqn}. The values of $\Sigma_{\rm HI}$ are rough estimates, obtained by scaling the observed optical radii. The error bars account for statistical uncertainties only.}
\label{fig_ks}
\end{figure*}

\subsubsection{Kennicutt-Schmidt Law}
\label{kslaw}

Although \hi~gas fraction and sSFR do not appear strongly correlated, gas surface density may be the more relevant parameter for star formation. According to the Kennicutt-Schmidt Law \citep{kennicutt98}, galaxies with a higher SFR surface density should have a higher gas surface density. We calculate $\Sigma_{\rm SFR}$~from Equation~\ref{sfi_eqn}. Due to the 3.5\arcmin~resolution of the Arecibo beam, we do not have radii for the \hi~gas. However, \hi~diameters are observed to scale with galaxy optical diameters \citep[\eg][]{broeils97, swaters02}. \citet{broeils97} find that \hi\ radii are typically 1.7 times larger than optical radii. We therefore use the SDSS radius containing 90$\%$~of the $r$-band light, $R_{90}$ to estimate the \hi~radius and calculate $\Sigma_{\rm HI}$~as
\begin{equation}
\label{hisd_eqn}
\Sigma_{\rm HI}=\frac{M_{\rm HI}}{2 \pi (1.7 R_{90})^2}.
\end{equation} Since galaxies show substantial scatter in the ratios of their \hi\ and optical radii \citep{broeils97}, these $\Sigma_{\rm HI}$~values are only rough estimates.

We plot $\Sigma_{\rm SFR}$~and $\Sigma_{\rm HI}$~for the sample in Figure~\ref{fig_ks}. We find no clear correlation between the \hi~and SFR surface densities, as indicated by the low Spearman's rank correlation coefficient of 0.16. This lack of correlation is consistent with results that show that molecular gas primarily sets the Kennicutt-Schmidt Law \citep{schruba11}. Most galaxies lie slightly below the threshold density of $\sim$10 \Msol~pc$^{-2}$~for conversion to molecular gas. On average, the starbursts tend to have elevated $\Sigma_{\rm HI}$~relative to the rest of the sample; the high EW starbursts have an average $\Sigma_{\rm HI}$~more than 3 times higher than the non-starbursts, and the high sSFR starburst starbursts have an average $\Sigma_{\rm HI}$~2 times higher than the non-starbursts. Higher surface densities may aid the conversion of atomic gas to molecular gas and fuel star formation. Alternatively, the starbursts may have more spatially extended \hi~than the other galaxies, or they may have lower metallicities and higher \hi~saturation thresholds \citep[\eg][]{bolatto11}. As with \hi~gas fraction, the starbursts' $\Sigma_{\rm HI}$~values tend to be high, but they do not differ dramatically from the non-starbursts in the sample. 

Whether we consider sSFR, H$\alpha$~EW, extinction-corrected $g-r$~color, or $\Sigma_{\rm SFR}$, the \hi~content of galaxies shows only a weak connection with star formation. The starbursts have high \hi~gas fractions compared to the full ALFALFA~H$\alpha$~sample but show little to no increase in \hi~gas fraction relative to similar mass galaxies. These observations suggest that the \hi~content of gas-rich galaxies remains relatively constant, even during a starburst episode. Any excess \hi~is efficiently converted into \htwo, and photodissociation of \htwo~may balance the consumption or ionization of \hi.

\subsection{Star Formation Efficiency}
\label{sec_sfe}
\begin{figure*}
\epsscale{1.0}
\plotone{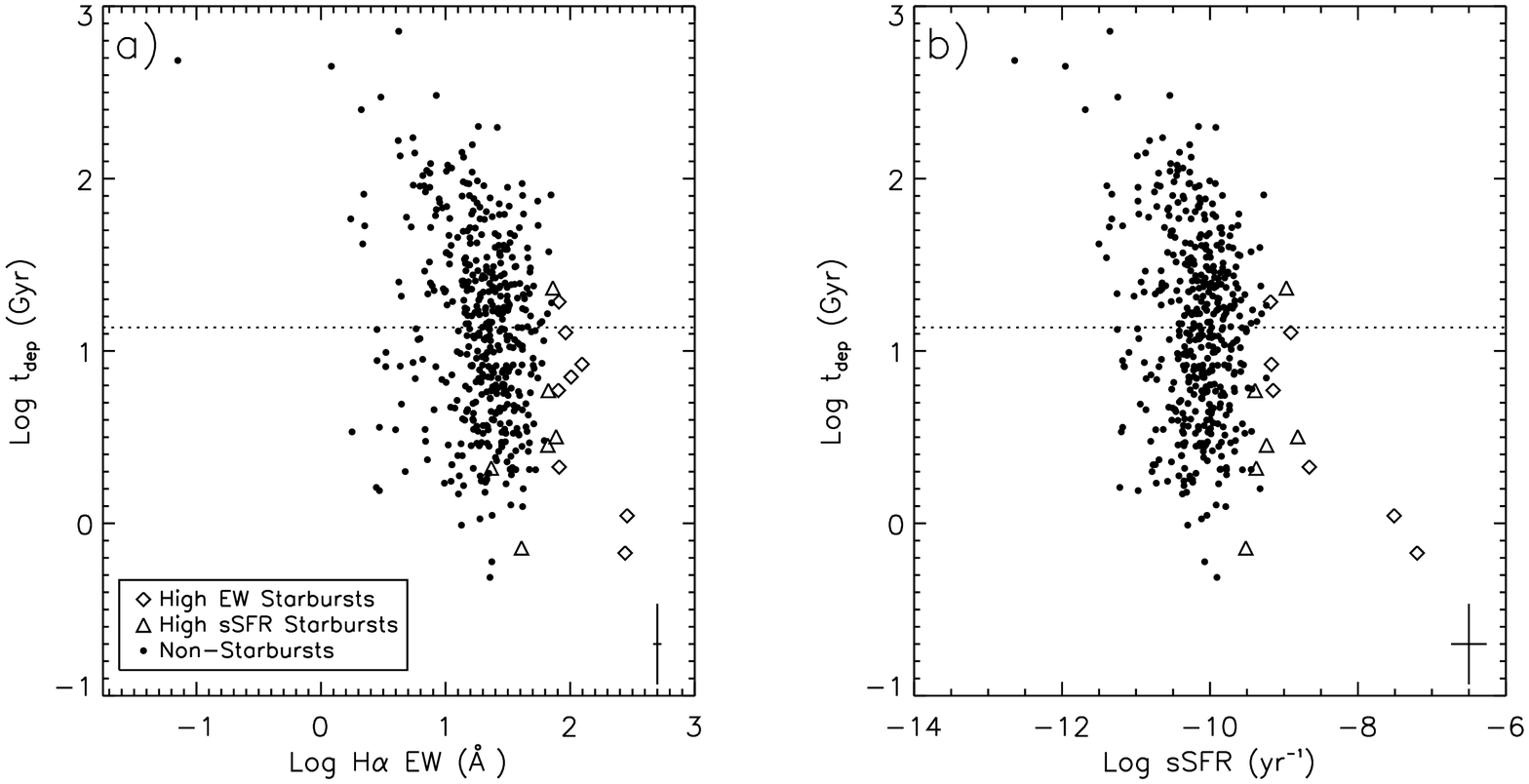}
\caption[(\emph {a}) The relation between $t_{\rm dep}$~and H$\alpha$~EW and (\emph {b}) the relation between $t_{\rm dep}$~and sSFR.]{The relation between $t_{\rm dep}$~and H$\alpha$~EW (left panel) and $t_{\rm dep}$~and sSFR (right panel). The right panel only includes galaxies with reliable SDSS photometry. The dashed line in both panels indicates the Hubble time. Despite their high \hi~content, most starbursts have shorter $t_{\rm dep}$~than the rest of the sample.}
\label{fig_gt_ssfr}
\end{figure*}

Since the starbursts have slightly high, but not unusual, \hi~content for their masses, their current SFRs imply a high \hi-to-\htwo\ conversion efficiency. Following\citet{schiminovich10}, we refer to the SFR per unit \hi\ gas mass as the \hi-based star formation efficiency (SFE). This parameter indicates how efficiently galaxies are able to tap into their \hi\ reservoirs and convert their \hi\ to molecular gas and stars. We calculate the inverse of the SFE, the \hi~gas depletion timescale, as
\begin{equation}
\label{sfe_eqn}
t_{\rm dep}=\frac{M_{\rm HI}}{\rm SFR},
\end{equation} and we show the relation between $t_{\rm dep}$~and sSFR in Figure~\ref{fig_gt_ssfr}. The starburst \hi~depletion times are listed in Table~\ref{sbtable}. Most of the sample shows no correlation between sSFR or EW and \hi~depletion time. However, the starbursts tend to have short \hi~depletion times, despite the fact that they are \hi-rich. All the high EW starbursts and four out of six high sSFR starbursts have \hi~gas fractions above the sample median, and only two starbursts have depletion times below the sample median. In particular, the two strongest or youngest starbursts, as measured by EW, have both the highest \hi~gas fractions and among the highest SFE. \citet{saintonge11b} suggest that weak starbursts in the high-mass GASS sample may not be able to access \hi~easily, since they have the same \hi~$t_{\rm dep}$~as non-starbursts. In contrast, our starbursts do show lower $t_{\rm dep}$, which may indicate that lower mass starbursts are efficiently converting \hi~to \htwo~and stars.

\begin{figure*}
\epsscale{1.0}
\plotone{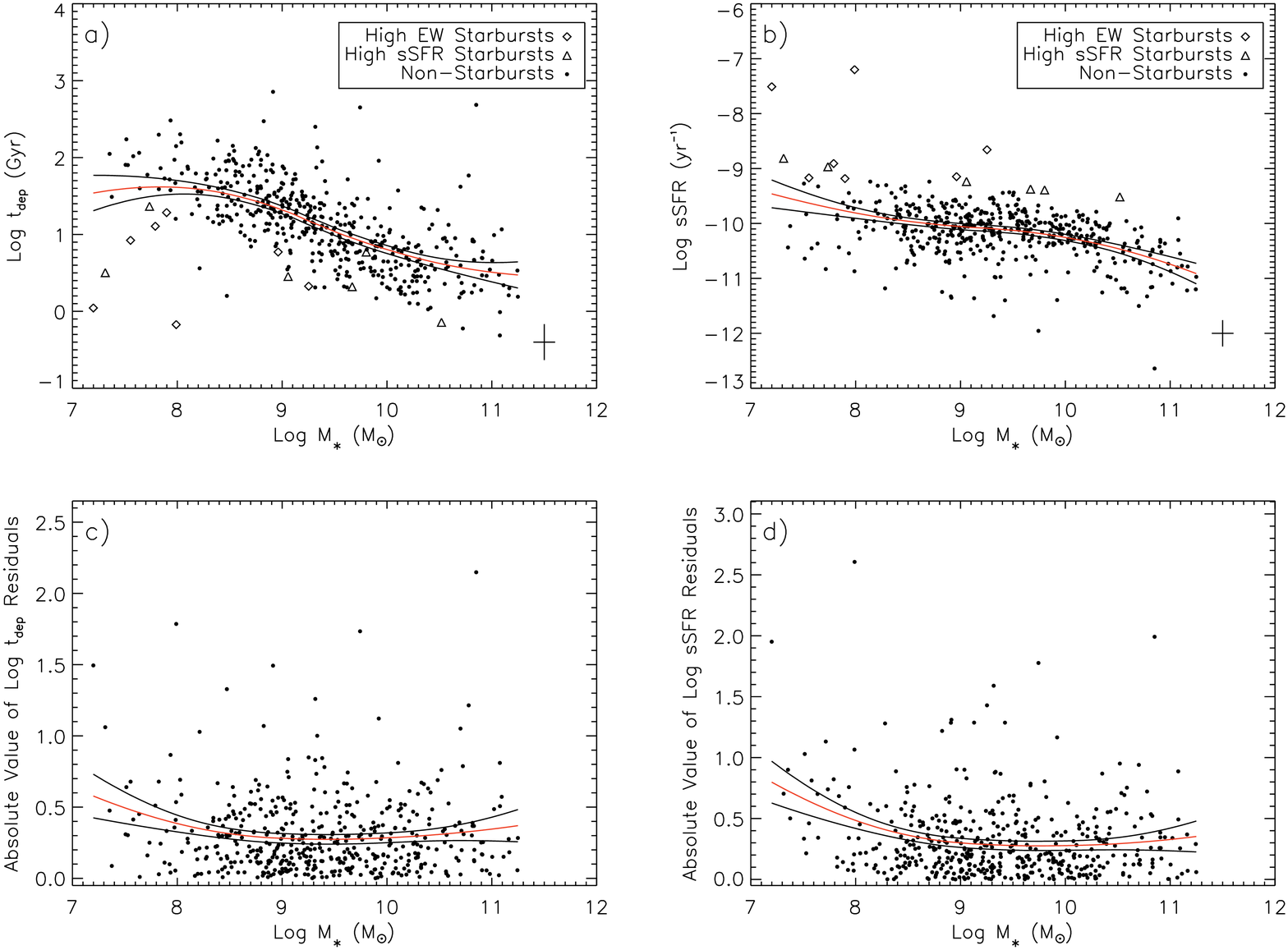}
\caption[(\emph {a}) $t_{\rm dep}$~and (\emph {b}) sSFR as a function of $M_*$. ]{(\emph {a}) $t_{\rm dep}$~and (\emph {b}) sSFR as a function of $M_*$. The red line shows the mean value at each $M_*$, as determined by a LOESS non-parametric local regression fit. The black lines show the 95\% confidence intervals of the mean. In panels (\emph {c}) and  (\emph {d}) we show the absolute value of the residuals about the mean fits in the upper panels. The red line shows a LOESS fit to the residuals, and the black lines again show the 95\% confidence intervals. Low mass galaxies show higher scatter about the mean for both $t_{\rm dep}$ and sSFR.}
\label{fig_gtbins}
\end{figure*}

In general, $t_{\rm dep}$~decreases slightly with $M_*$ (Figure~\ref{fig_gtbins}a), for masses below 10$^{10}$\Msol\ \citep[\eg][]{bothwell09,huang12}. The high \hi~gas fractions and low SFE observed for the lowest mass non-starburst galaxies may imply that their accretion rates exceed their rates of gas consumption \citep[\eg][]{huang12, kannappan13}, either due to inefficient inward gas transport or inefficient \hi-to-\htwo\ conversion. However, the low-mass end also contains several galaxies with $t_{\rm dep}$~an order of magnitude lower than the average for their stellar mass. In addition to their higher scatter in $t_{\rm dep}$, the lowest mass galaxies also appear to exhibit the largest scatter in sSFR (Figure~\ref{fig_gtbins}b). \citet{lee07}~explain the scatter in sSFR as an increase in``burstiness" at low stellar masses.

We examine this change in scatter quantitatively in Figure~\ref{fig_gtbins}. In Figures~\ref{fig_gtbins}a and \ref{fig_gtbins}b, we fit the mean $t_{\rm dep}$ and mean sSFR as a function of $M_*$ using the LOESS non-parametric local regression technique \citep{cleveland79} in R \citep{r14}. The LOESS method determines the mean at each $M_*$ by weighting data points based on their distance from the $M_*$ value of interest. To illustrate the scatter about this mean, we plot the absolute value of the residuals from the mean fit in Figures~\ref{fig_gtbins}c and \ref{fig_gtbins}d. A LOESS fit to the residuals supports our rough by-eye assessment that the scatter about the mean in both $t_{\rm dep}$ and sSFR is highest among the lowest mass galaxies. The gas-richness of star-forming dwarf galaxies, combined with their wide-ranging SFE, suggest that \hi~gas may accumulate until a dynamical disturbance triggers gas inflows and compression.

\begin{figure*}
\epsscale{1.0}
\plotone{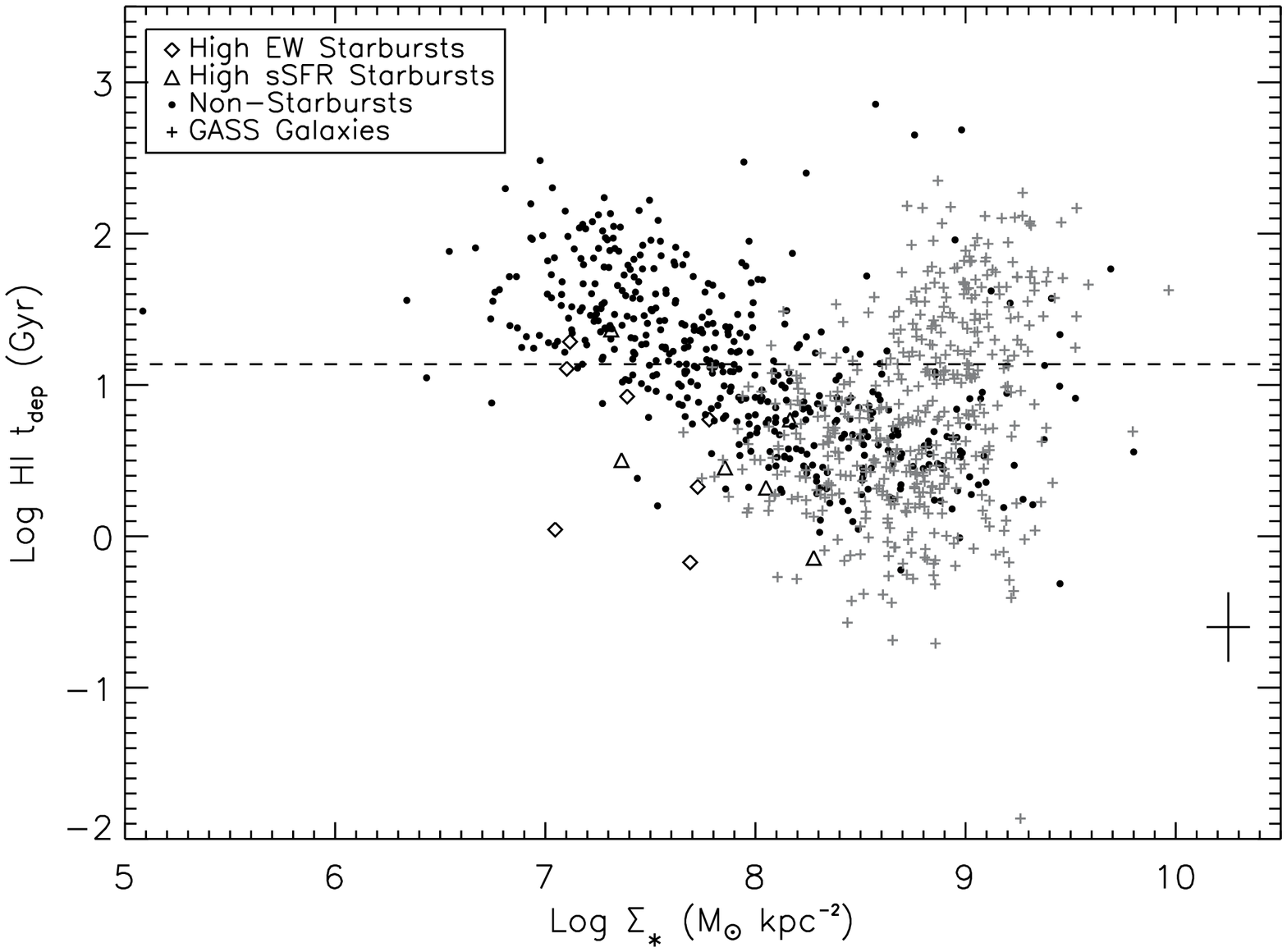}
\caption[$t_{\rm dep}$~as a function of $\Sigma_*$~for the ALFALFA~H$\alpha$~sample and the GASS sample.]{$t_{\rm dep}$~as a function of $\Sigma_*$~for the ALFALFA~H$\alpha$~sample (black circles, diamonds, and triangles) and the GASS sample (gray crosses). Below the ``quenching threshold" of $\Sigma_*=10^{8.7}$ \Msol kpc$^{-2}$ \citep{catinella10,saintonge11a}, $t_{\rm dep}$~decreases with increasing $\Sigma_*$. The black cross indicates the median statistical errors for the ALFALFA~H$\alpha$~sample, and the dashed line indicates the Hubble time.}
\label{fig_ssd_gt}
\end{figure*}

Galaxy structure may also play a role in determining \hi~to \htwo~conversion efficiency and setting the \hi~$t_{\rm dep}$. \citet{blitz06}~argue that in disk galaxies, the midplane pressure sets the \htwo/\hi~ratio and hence the SFE. Since the midplane pressure should scale with stellar mass surface density, $\Sigma_*$~\citep[\eg][]{blitz04}, we plot $t_{\rm dep}$~as a function of $\Sigma_*$~in Figure~\ref{fig_ssd_gt}. We calculate 
\begin{equation}
\label{ssd_eqn}
\Sigma_*=\frac{M_*}{2 \pi R_{50,r}^{2}}
\end{equation} where $R_{50,r}$~is the SDSS $r$-band \citet{petrosian76} half-light radius. Figure~\ref{fig_ssd_gt}~shows that for the ALFALFA~H$\alpha$~galaxies, $t_{\rm dep}$~decreases with $\Sigma_*$, as expected if higher surface density disks convert \hi~to stars more efficiently.  

However, this trend is the exact opposite of that found by \citet{saintonge12}~for high-mass, optically-selected GASS galaxies, where the mean \hi~$t_{\rm dep}$~increases with $\Sigma_*$. We show the \hi~$t_{\rm dep}$~and $\Sigma_*$ of the GASS sample \citep{catinella10,catinella12,catinella13}, calculated following Equations~\ref{sfe_eqn}~and \ref{ssd_eqn}, in Figure~\ref{fig_ssd_gt}. We only include GASS galaxies with \hi\ detections, and we use the ``representative" GASS sample, which includes gas-rich galaxies \citep{catinella13}. The GASS stellar masses are calculated from SDSS photometry \citep{catinella13}. We calculate SFRs from the tabulated {\it GALEX} NUV photometry as described in \citet{schiminovich10}, assuming a \citet{chabrier03} IMF. Among massive galaxies, UV-based SFRs agree well with H$\alpha$-based estimates \citep[\eg][]{salim07}, which therefore permits a direct comparison with our sample.

The discrepancy between the ALFALFA~H$\alpha$~galaxies and the GASS galaxies is likely due to the quenching of star formation in more massive galaxies. Figure~\ref{fig_ssd_gt} shows that the scatter in $t_{\rm dep}$~increases dramatically above a $\Sigma_*$~of 10$^{8.7}$ \Msol~kpc$^{-2}$, identified as a ``quenching threshold" by \citet{catinella10} and \citet{saintonge11a}. The \hi~depletion times of the few ALFALFA~H$\alpha$~galaxies in this high stellar surface density regime are consistent with those of the GASS population. \citet{saintonge12} suggest that galaxies with high stellar surface densities may be more stable to fragmentation due to their bulge-dominated morphologies \citep[\eg][]{ostriker73} or may no longer have access to their \hi~reservoir. 

\begin{figure*}
\epsscale{1.0}
\plotone{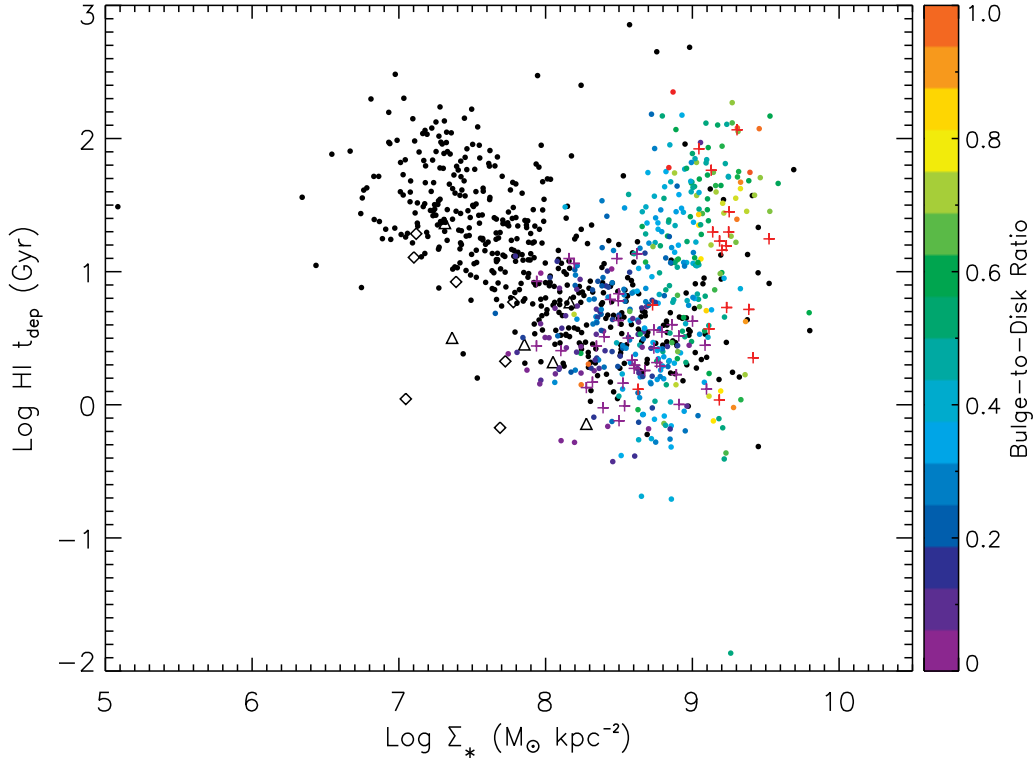}
\caption[$t_{\rm dep}$~as a function of $\Sigma_*$~for the ALFALFA~H$\alpha$~sample and the GASS sample, with galaxy morphologies indicated.]{$t_{\rm dep}$~as a function of $\Sigma_*$~for the ALFALFA~H$\alpha$~non-starbursts (black circles) and starbursts (diamonds and triangles) and the GASS sample (colored circles and crosses). The GASS galaxies are colored by bulge-to-disk ratio from the \citet{simard11}~SDSS fits. The crosses indicate GASS galaxies for which a pure S{\'e}rsic profile is preferable; red crosses indicate a S{\'e}rsic index $\ge 2$, and purple crosses indicate a S{\'e}rsic index $< 2$.}
\label{fig_gt_morph}
\end{figure*}

We use the SDSS surface brightness profile fits of \citet{simard11} to quantify the morphologies of the GASS galaxies. These fits are not available for the ALFALFA~H$\alpha$~galaxies, which fall in the more recent, SDSS DR9 sky coverage. Following \citet{simard11}, we adopt a bulge$+$disk model if the probability that the bulge$+$disk model is not required is below 0.32; if the probability is higher, we use a pure S{\'e}rsic model. Figure~\ref{fig_gt_morph}~shows that the GASS disk galaxies continue the trend of decreasing $t_{\rm dep}$~with $\Sigma_*$~seen in the ALFALFA~H$\alpha$~sample, while spheroid-dominated galaxies deviate to higher $t_{\rm dep}$. Although all the offset galaxies appear to be bulge-dominated, they show no discernible trend between bulge-to-disk ratio and $t_{\rm dep}$. 

\begin{figure*}
\epsscale{0.6}
\plotone{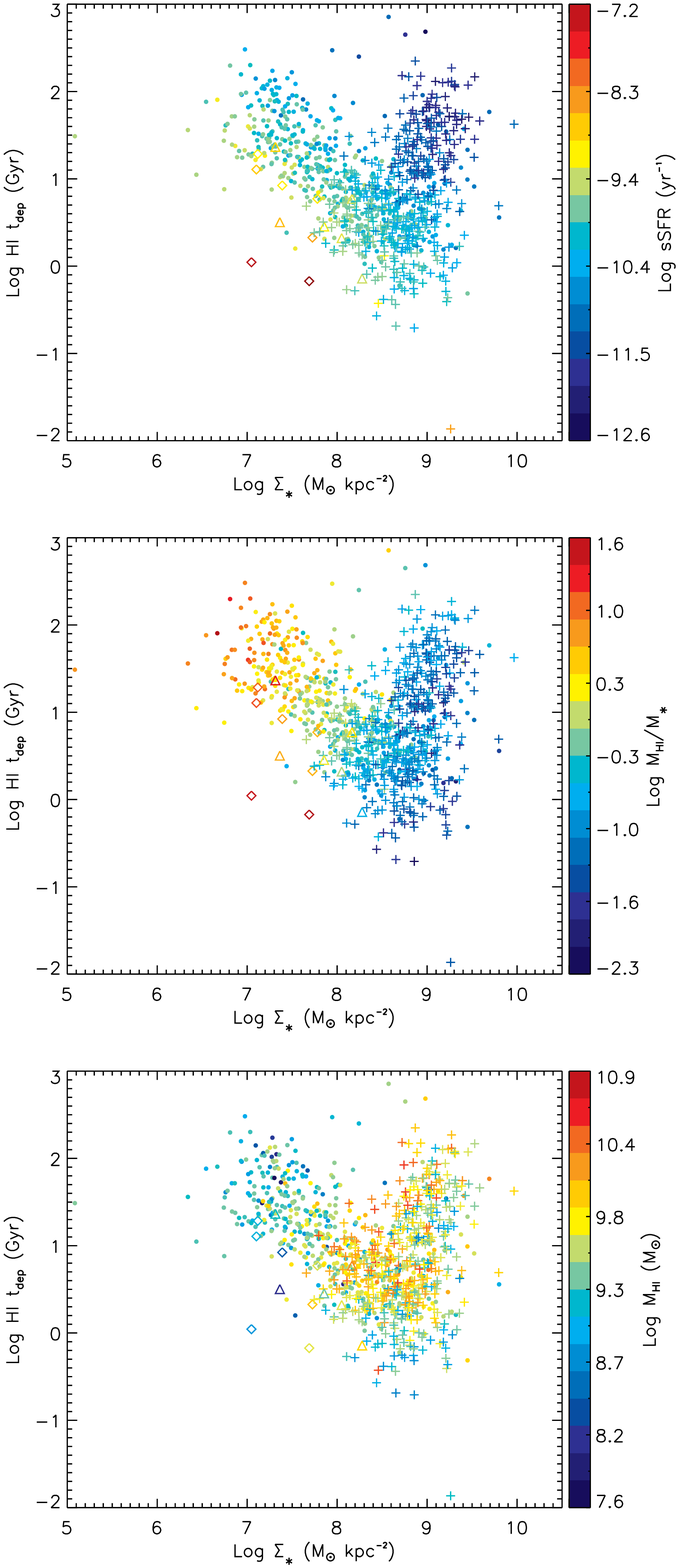}
\caption[$t_{\rm dep}$~as a function of $\Sigma_*$~with (\emph {a}) sSFR, (\emph {b}) $M_{\rm HI}$/$M_*$, and (\emph {c}) $M_{\rm HI}$~indicated.]{$t_{\rm dep}$~as a function of $\Sigma_*$~for the ALFALFA~H$\alpha$~non-starbursts (circles) and starbursts (diamonds and triangles) and the GASS sample (crosses). Color indicates sSFR (top panel), \gf~(center panel), and \mhi~(bottom panel). }
\label{fig_gt_gas}
\end{figure*}

The spheroidal galaxies also appear offset to lower sSFRs (Figure~\ref{fig_gt_gas}a) compared to the rest of the GASS and ALFALFA~H$\alpha$~galaxies, supporting the idea that they have quenched their star formation.  Interestingly, at a given $\Sigma_*$, the \gf~values of spheroidal galaxies are independent of sSFR (Figure~\ref{fig_gt_gas}b), indicating that the dispersion in $t_{\rm dep}$~at the high $\Sigma_*$~end is not the result of a wide range in \hi~content. Instead, the low sSFR galaxies appear unable to use their \hi~efficiently. In terms of absolute $M_{\rm HI}$, galaxies with the highest \hi~masses tend to live just below the quenching threshold  (Figure~\ref{fig_gt_gas}c); galaxies with higher $\Sigma_*$~must have lost or consumed their \hi~gas during the quenching process. The lack of a trend with bulge-to-disk ratio or \gf~among the quenched galaxies suggests that differences in the spatial distribution of the \hi~gas may account for their variation in SFE.

While the spheroidal galaxies exhibit the highest $t_{\rm dep}$~for their $\Sigma_*$, starbursts have low $t_{\rm dep}$~relative to galaxies of a similar $M_*$~or $\Sigma_*$ (Figures~\ref{fig_gtbins}-\ref{fig_gt_gas}). Many studies have pointed out the role mergers may play in enhancing SFE \citep[\eg][]{solomon88,combes94,bouche07,dimatteo07,bournaud11}. Turbulence and gas flows during major mergers can lead to the efficient formation of molecular gas, shifting the ISM gas distribution to higher densities \citep{bournaud11,powell13}. During a gas-rich merger, the total gas fraction of a galaxy may increase due to the addition of new gas; however, if the distribution of gas becomes skewed to higher gas densities, the \htwo~gas fraction may increase more strongly than the \hi~gas fraction. This scenario could potentially explain the moderately high \hi~gas fractions and short $t_{\rm dep}$~of the ALFALFA~H$\alpha$~starbursts.

\subsection{Morphology and Mergers}
\label{sec_merge}

The starbursts' morphologies and kinematics may demonstrate whether dynamical disturbances are enhancing the SFE. One measure of morphological disturbance is the 180$^{\circ}$ rotational asymmetry. We calculate the $R$-band asymmetry, $A_R$, following \citet{conselice00a}, as 
\begin{equation}
\label{asym_eqn}
A_R={\rm min}(\frac{\sum \left | I_0-I_{180} \right |}{\sum \left |I_0 \right |})-{\rm min}(\frac{\sum \left | B_0-B_{180} \right |}{\sum \left |I_0 \right |}),
\end{equation} where $I_0$ is the original image, $I_{180}$ is the image rotated by 180$^{\circ}$, $B_0$ is a background region, and $B_{180}$ is the background region rotated by 180$^{\circ}$. We sum over all the image pixels within the Petrosian radius \citep{petrosian76} at the $\eta=0.2$ surface brightness level, where $\eta$ is the ratio of the intensity at a given radius to the average intensity within that radius. We also scale the background region area to this radius. The minimization finds the rotational center that produces the lowest asymmetry value. Prior to the calculation, all images are background-subtracted, and we mask all stars within approximately five galaxy radii using the IRAF\footnote{IRAF is distributed by the National Optical Astronomy Observatories, which are operated by the Association of Universities for Research in Astronomy, Inc., under cooperative agreement with the National Science Foundation.} task {\tt imedit}. As recommended by \citet{conselice00a}, we adopt a signal-to-noise (S/N) cut of 100. Below this value, the scatter in $A_R$ dramatically increases, and 41\% of these low S/N ALFALFA~H$\alpha$ galaxies have unphysical negative $A_R$ values. 

\begin{figure*}
\epsscale{1.0}
\plotone{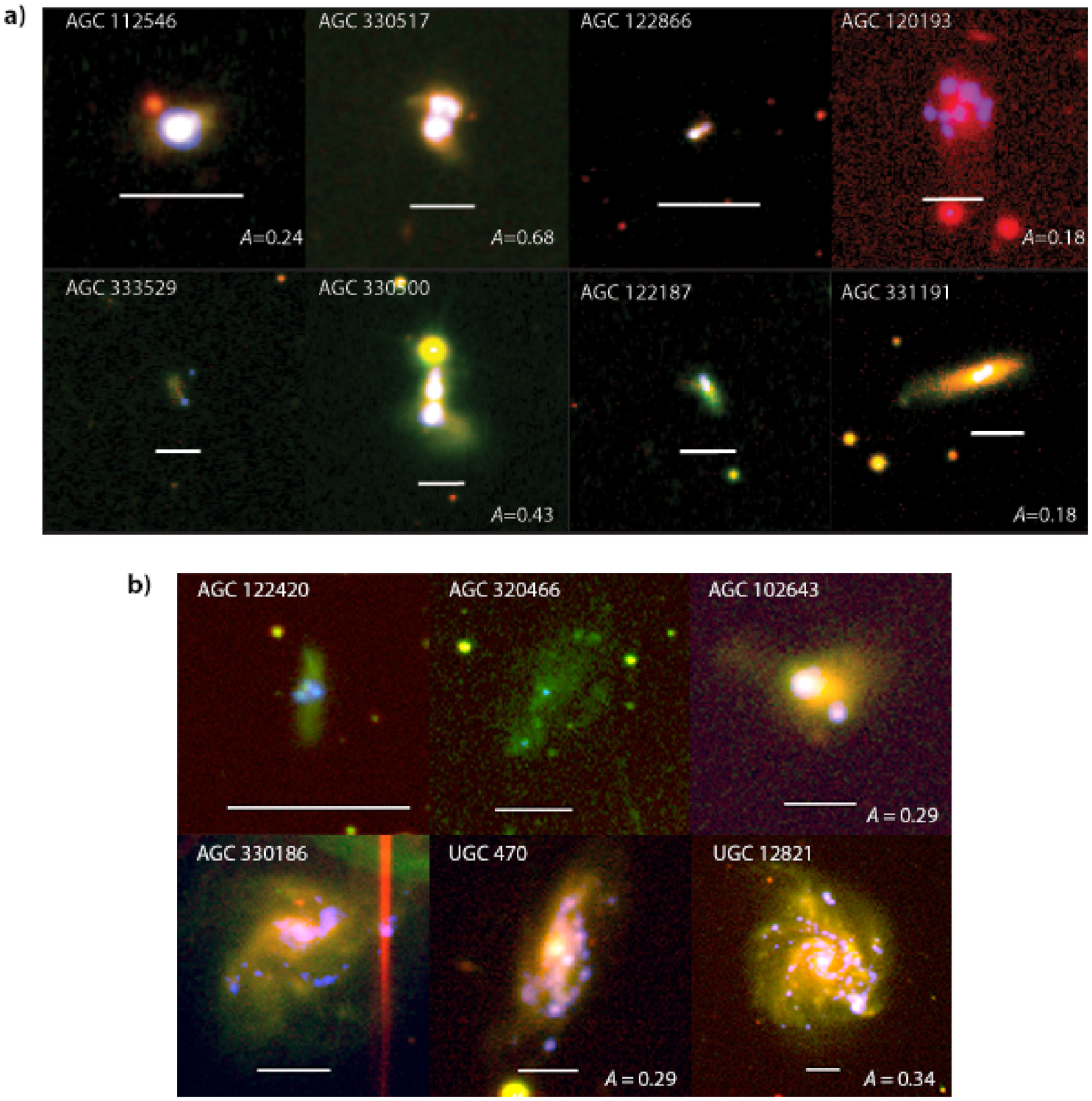}
\caption[Morphologies of the eight ALFALFA~H$\alpha$~starbursts.]{Morphologies of the eight high EW ALFALFA~H$\alpha$~starbursts (top panel), organized in order of H$\alpha$~EW (see Table~\ref{sbtable}). The highest EW starburst is at the upper left, and the lowest EW starburst is at the lower right. The bottom panel shows the morphologies of the six high sSFR starbursts organized by decreasing sSFR. The ALFALFA~H$\alpha$ $R$-band images appear in red, SDSS $g$-band images appear in green, and the continuum-subtracted H$\alpha$~emission is shown in blue. All images have a logarithmic brightness scale. AGC 120193 does not have SDSS images. The solid white line in each panel corresponds to 5 kpc. The $R$-band asymmetries are indicated for the eight starbursts with S/N$>100$. The yellow region above AGC 330500 is a bright star.}
\label{fig_sbimages}
\end{figure*}

\begin{figure*}
\epsscale{0.7}
\plotone{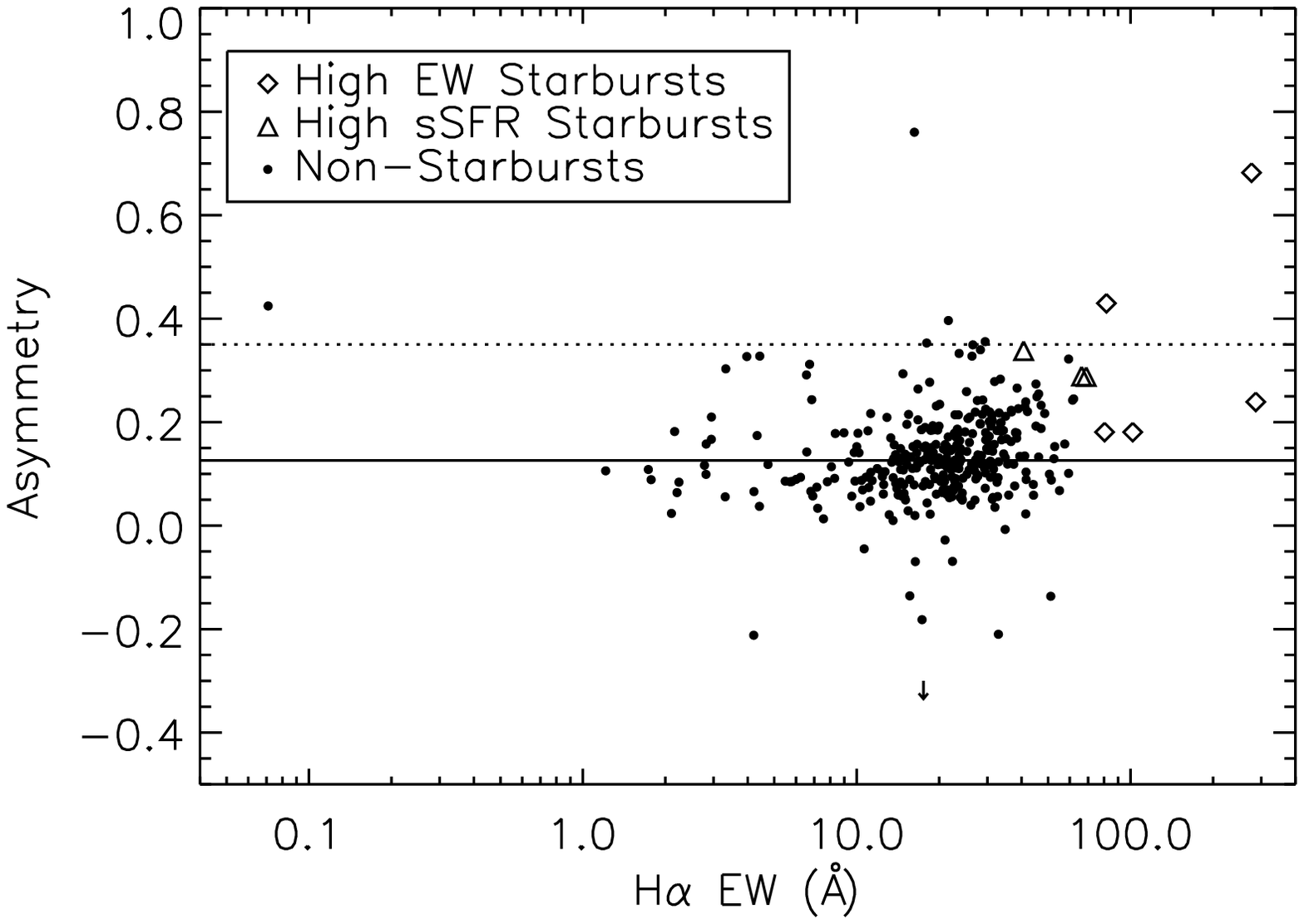}
\caption[The $R$-band asymmetries and H$\alpha$~EWs for ALFALFA~H$\alpha$~galaxies.]{The $R$-band asymmetries and H$\alpha$~EWs for galaxies with S/N$>100$. The solid line shows the median asymmetry and the dotted line shows the $A_R>0.35$ criterion for identifying major mergers. The downward arrow indicates an outlier with unphysical negative asymmetry. For clarity, two overlapping data points for the high sSFR starbursts have been artificially displaced in EW.}
\label{fig_asym_ew}
\end{figure*}

We show images of the eight high EW and six high sSFR ALFALFA~H$\alpha$~starbursts and their calculated asymmetries in Figure~\ref{fig_sbimages} and list the asymmetry values in Table~\ref{sbtable}. The eight starbursts with S/N $>100$ have asymmetries ranging from 0.18-0.68, higher than the median sample asymmetry of 0.14. \citet{conselice03} suggests that asymmetries above $\sim$0.35 indicate major mergers. Figure~\ref{fig_asym_ew} shows the asymmetries of the sample as a function of H$\alpha$ EW. All the starbursts have slightly elevated asymmetries, and two (AGC 330517 and AGC 330500) are clearly major mergers. However, we find equally disturbed galaxies at lower values of H$\alpha$ EW, indicating that not all \hi-rich mergers are starbursts.

If some of the starbursts are merging systems, they appear to be near coalescence. Simulations of major mergers also indicate that the peak star formation activity should occur at this time \citep[\eg][]{mihos94,cox08,lotz10a}. Intriguingly, the two most asymmetric starbursts, AGC 330517 and AGC 330500, are also the high EW starbursts with the highest offsets from the \hi~gas fraction and stellar mass relation (see Figure~\ref{fig_gf_mstar}).  High gas-fraction mergers experience more extended star formation, which causes them to appear asymmetric over a longer timescale \citep{lotz10b}. Alternatively, these two starbursts may be in an earlier merger stage than the other starbursts, prior to coalescence, when they have not yet consumed or expelled much of their \hi~gas. The individual merging galaxies in AGC 330517 and AGC 330500 are still distinguishable in Figure~\ref{fig_sbimages}, which supports this possibility. The other six starbursts with asymmetry measurements have $A_R\sim0.2-0.3$, only slightly higher than the average asymmetry of the full sample. These lower asymmetries could simply reflect clumpy star formation in these galaxies \citep[\eg][]{conselice00a,fossati13}, as appears to be the case for UGC 12821, or they could indicate a mild dynamical disturbance. The three most massive high sSFR starbursts, AGC 330186, UGC 470, and UGC 12821, show spiral structure and no signs of a major interaction. However, the presence of a nearby bright star blocks a full view of AGC 330186, and the galaxy does appear somewhat asymmetric in its structure. Minor disturbances may cause the enhanced star formation in these spirals. AGC 120193, AGC 331191, and AGC 102643 show more obvious signs of morphological disturbance, with extended, asymmetric tails visible in their broadband images. The lower asymmetries in some of the starbursts could also be consistent with a later merger stage; the simulations of \citet{lotz10a} show that the peak star formation in major mergers may occur a few hundred Myr after the peak asymmetry. Mild asymmetries of $A_R\sim0.2$ correspond to this peak star-forming, coalescence stage in mergers. If these less asymmetric high EW starbursts are indeed late-stage mergers, their lower \hi\ gas fractions relative to the more asymmetric starbursts AGC 330517 and AGC 330500 suggest that dynamical disturbances trigger the efficient conversion of \hi~into \htwo~throughout the merger. The more asymmetric, earlier stage mergers, such as AGC 330517 and AGC 330500, still exhibit excess \hi, which may be consumed or ionized by the end of the starburst phase.

\begin{figure*}
\epsscale{0.7}
\plotone{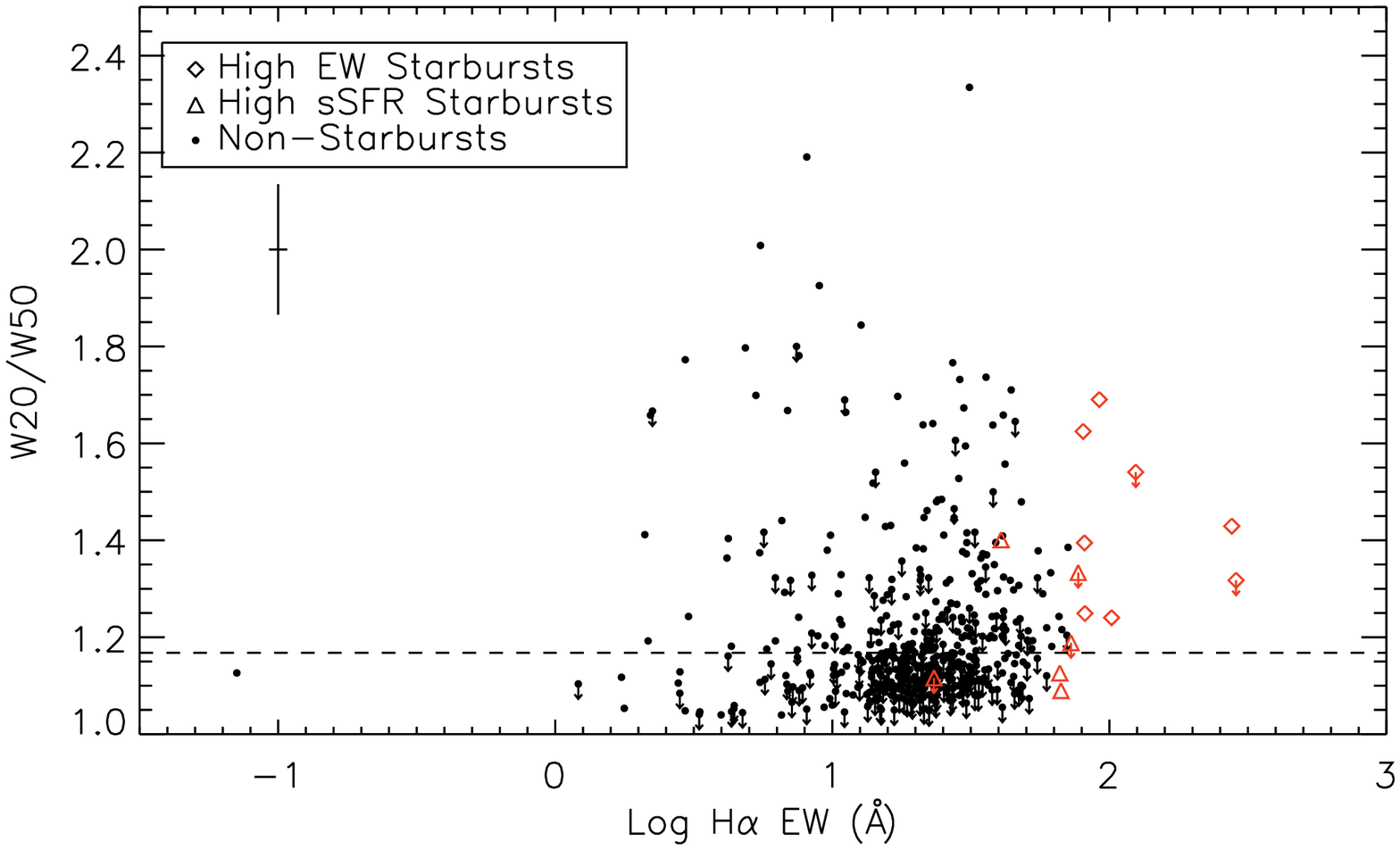}
\caption[\hi~velocity width ratios, $W_{20}/W_{50}$, and H$\alpha$~EWs of the ALFALFA~H$\alpha$~sample.]{\hi~velocity width ratios, $W_{20}/W_{50}$, and H$\alpha$~EWs of the ALFALFA~H$\alpha$~sample. The dashed line shows the median $W_{20}/W_{50}$; this median includes galaxies with upper limits on $W_{20}/W_{50}$, and therefore may be overestimated. The starbursts, shown by red diamonds and triangles, tend to have above average \hi~width ratios. Representative error bars are shown at the upper left of the plot.}
\label{fig_w2050_ew}
\end{figure*}

Disturbed gas kinematics may also be a sign of galaxy interactions. The \hi~velocity profiles of merging galaxies often exhibit asymmetries or wide, high-velocity wings due to an excess of high-velocity gas \citep[\eg][]{gallagher81,mirabel88}. To quantify this high-velocity excess, we calculate the ratio of the \hi~velocity width at 20\%~of the peak flux to the width at 50\% of the peak, $W_{20}/W_{50}$, as suggested by \citet{conselice00b}. Since the ALFALFA velocity resolution is 10 \kmps, we calculate an upper limit to $W_{20}/W_{50}$ for galaxies with $W_{20}/2-W_{50}/2<10$ \kmps. We show the \hi~width ratios and H$\alpha$~EWs of the sample in Figure~\ref{fig_w2050_ew}~and Table~\ref{sbtable}. All of the high EW starbursts have \hi~width ratios greater than the sample median, although as with morphological asymmetry, high \hi~width ratios occur in non-starbursts as well. In simulations of merging galaxies, \citet{powell13} find that interaction-driven turbulence, rather than supernova feedback, creates the high gas velocity dispersions observed in mergers. They argue that this enhanced turbulence increases the SFR by shifting the gas distribution to higher densities. The observed \hi~gas disturbances are therefore a possible cause of the high star formation efficiencies in the high EW starbursts.

\begin{figure*}
\epsscale{1.0}
\plotone{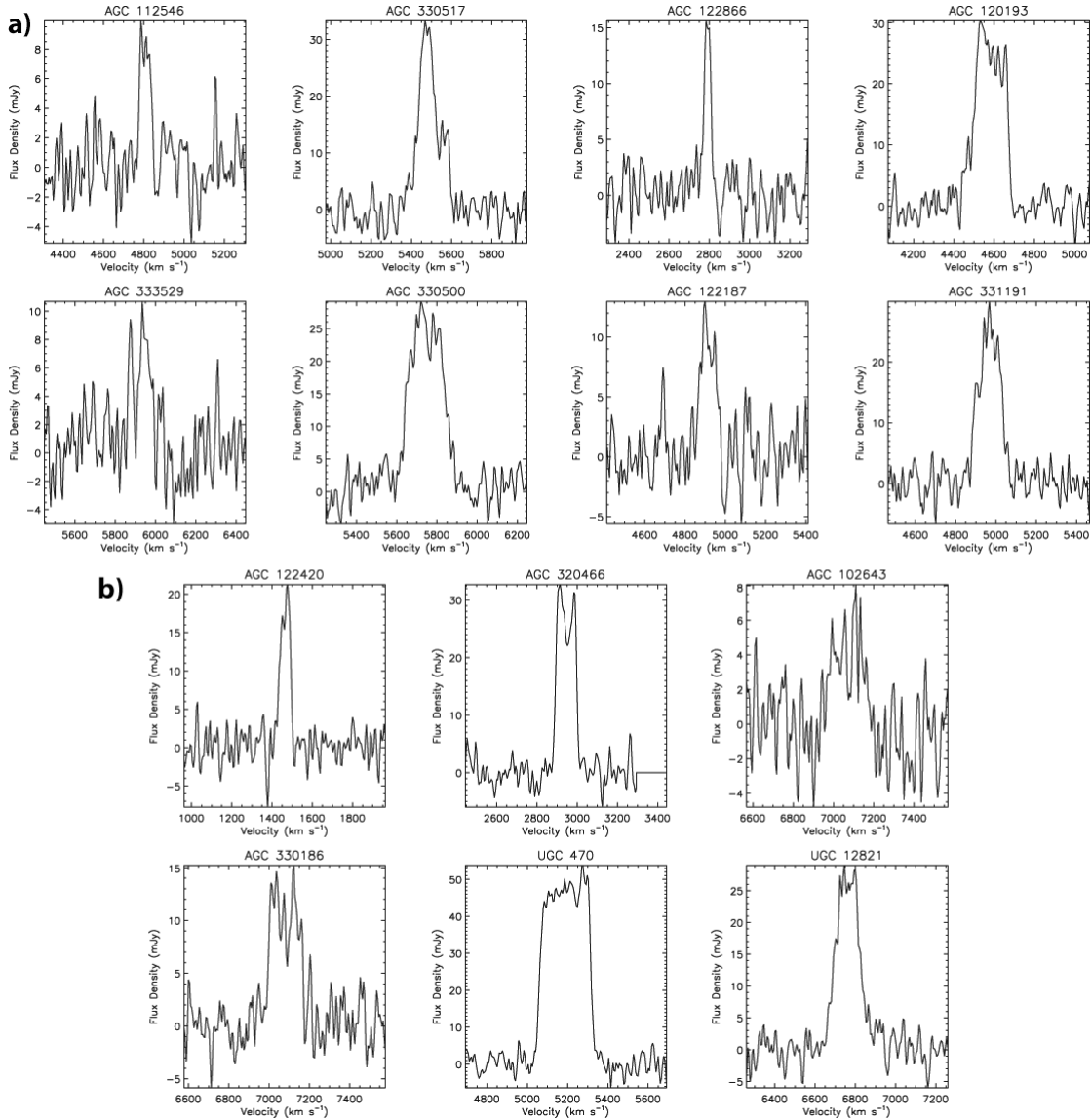}
\caption[The \hi~velocity profiles of the starbursts.]{The \hi~velocity profiles of the high EW starbursts (\emph {a}) and high sSFR starbursts (\emph {b}), ordered as in Figure~\ref{fig_sbimages}.}
\label{fig_sbvel}
\end{figure*}

\begin{figure*}
\epsscale{1.0}
\plotone{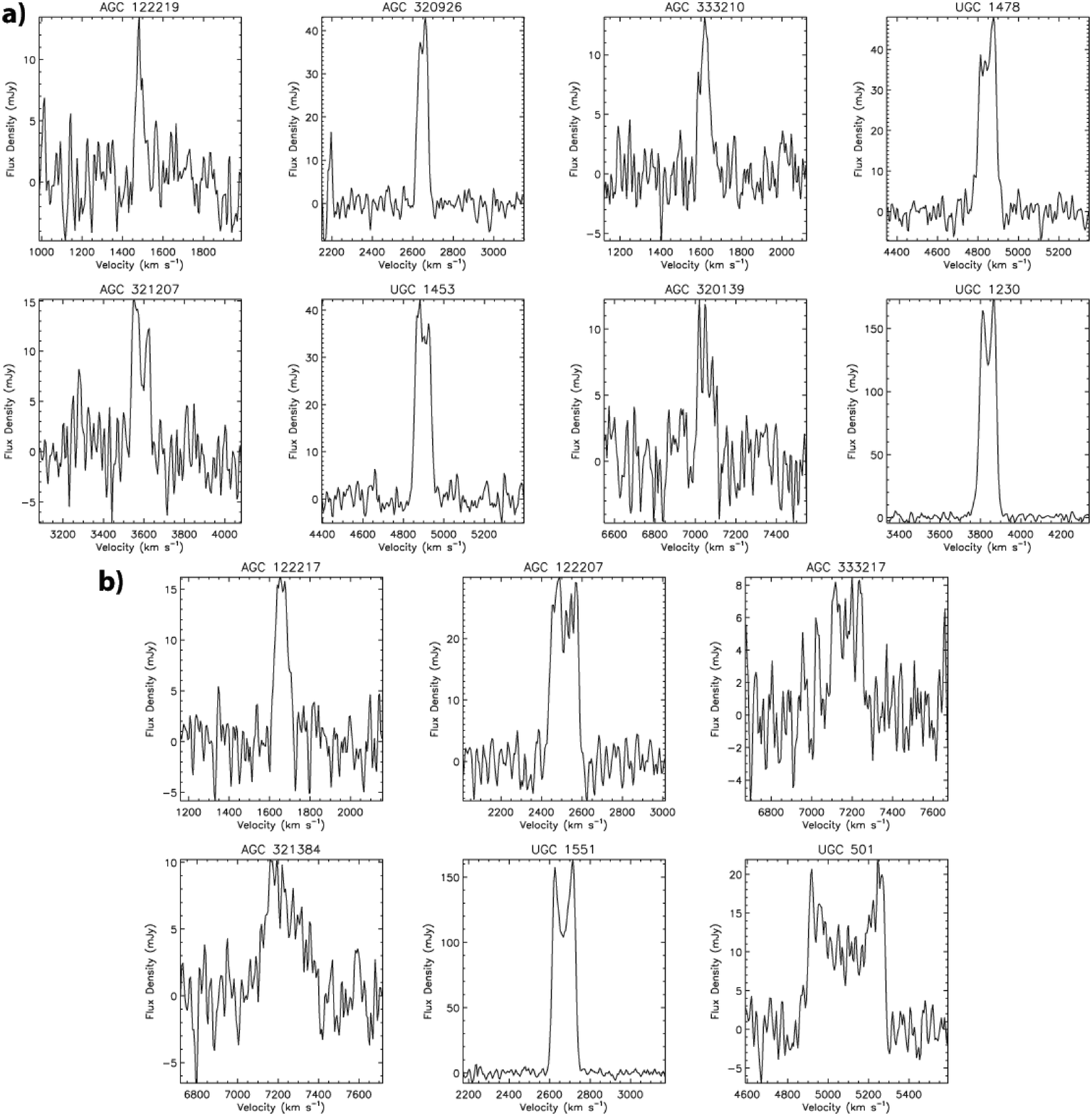}
\caption[The \hi~velocity profiles of eight non-starbursts.]{The \hi~velocity profiles of 14 non-starbursts, selected to have similar stellar masses and S/N spectra as the corresponding starbursts in Figure~\ref{fig_sbvel}. Since one starburst (AGC 120193) does not have a stellar mass estimate, we select the corresponding galaxy (UGC 1478) on the basis of \hi~mass.}
\label{fig_twins}
\end{figure*}

Non-interacting dwarf galaxies may also show broad wings in their \hi~profiles, however, and the high $W_{20}/W_{50}$~ratios of the starbursts do not necessarily prove they are interacting \citep[\eg][]{gallagher81}. We display the \hi~velocity profiles of the high EW and high sSFR starbursts in Figure~\ref{fig_sbvel}. While the \hi~profiles of several starbursts appear broad or irregular, the profiles of some of the lower-mass starbursts, especially AGC 112546 and AGC 122866, appear quite narrow and only have upper limits on $W_{20}/W_{50}$.  For comparison, we show examples of non-starburst \hi~profiles in Figure~\ref{fig_twins}. For each starburst, we randomly selected a non-starburst counterpart with $M_*$~within $\sim$0.5 dex and with a similar S/N \hi~spectrum. Since one starburst, AGC 120193, does not have SDSS photometry, we selected a non-starburst with $M_{\rm HI}$ within 0.5 dex. The \hi~velocity profiles of the four most massive high EW starbursts, AGC 330517, AGC 120193, AGC 330500, and AGC 331191, do appear to have broader wings than the straight-sided profiles of the non-starbursts. However, the profiles of the four lower mass high EW starbursts ($M_*<10^8$\Msol) are indistinguishable from non-starburst profiles of similar mass galaxies. We caution that the resolution of the \hi~spectra may be too low to accurately probe the profile shapes of these galaxies. Nevertheless, as discussed in \S~\ref{sec:hisbs}, the low-mass starbursts also have \hi~gas fractions comparable to those of low-mass non-starbursts. Therefore, the global \hi~gas fraction may not be the primary physical parameter that sets the level of star formation in low-mass galaxies. Infalling gas clouds are one possible trigger of star formation in dwarf galaxies \citep[\eg][]{gordon81,lopezsanchez12,verbeke14}. \citet{verbeke14} show that this type of interaction may generate multiple starburst episodes; disturbed gas kinematics may only be observable during the initial starburst and with favorable viewing orientations. These low-mass starbursts may therefore no longer show \hi~disturbances and may have consumed some of their \hi~during a prior burst.

The high sSFR starbursts generally show more ordered \hi\ velocity profiles than the high EW starbursts. As with the some of the high EW starbursts, two galaxies (AGC 122420 and AGC 102643) have inconclusive \hi\ profiles due to low resolution or low S/N. The other high sSFR starbursts show no obvious disturbances. The more massive spirals AGC 330186 and UGC 470 have straight-sided profiles indicative of orderly rotation, while the Gaussian profile of UGC 12821 is likely due to its face-on orientation \citep[\eg][]{haynes98}. Finally, the \hi-rich, low surface brightness galaxy AGC 320466 also shows ordered rotation in its \hi\ gas. With a high \mhi/$M_*=25$, much of AGC 320466's \hi\ gas may not reside near its star-forming regions. Unlike the broad profiles of several of the high EW starbursts, the ordered velocity profiles of most of the high sSFR starbursts do not indicate any major disturbances.

\begin{figure*}
\epsscale{1.0}
\plotone{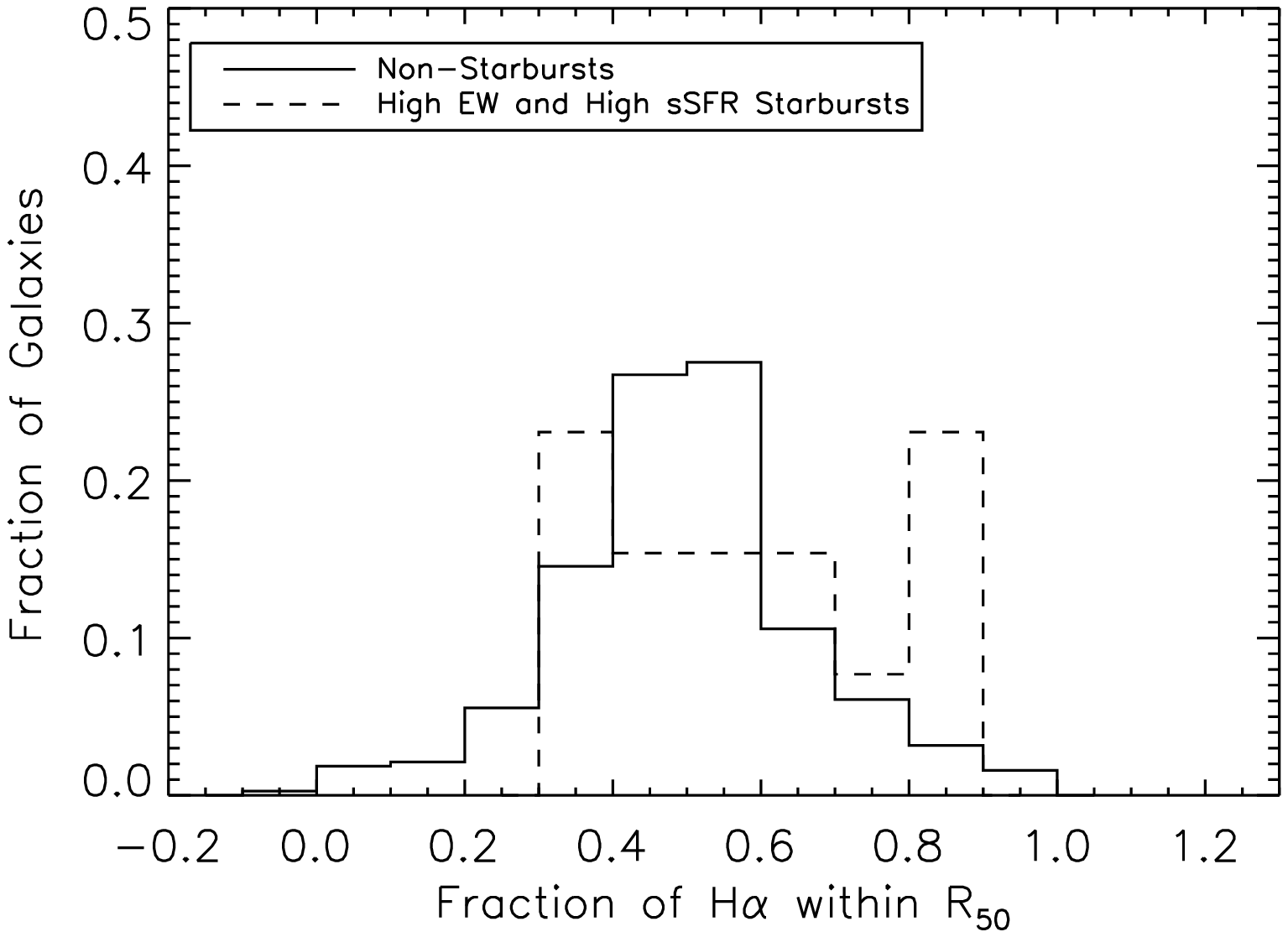}
\caption[The fraction of H$\alpha$~emission within the $R$-band half-light radius for non-starbursts  and starbursts.]{The fraction of H$\alpha$~emission within the $R$-band half-light radius for non-starbursts (solid lines) and starbursts (dashed lines). Only galaxies with an H$\alpha$~S/N$>10$~are shown. }
\label{fig_conc}
\end{figure*}

Although the traditional picture of merger-driven star formation emphasizes nuclear starbursts, recent studies have suggested that mergers may also cause a substantial increase in extended star formation \citep[\eg][]{ellison13,powell13}. In Figure~\ref{fig_conc}, we show the fraction of star formation within $R_{50}$~for starbursts and non-starbursts. We only include galaxies with a S/N$>10$ in the H$\alpha$~photometry. Three starbursts (AGC 112546, AGC 122866, and AGC 331191) show highly concentrated star formation, with more than 80\%~of their H$\alpha$~emission contained within $R_{50}$. However, other starbursts, particularly AGC 120193 and AGC 333529, show \hii~regions offset from the main $R$-band center of the galaxy. The majority of the star formation in these two starbursts is outside $R_{50}$. Gas flows in starbursts fuel nuclear star formation in many, but not all, cases.

\section{Discussion}
\label{sec:discussion}

\subsection{\hi~and Star Formation}
The ALFALFA~H$\alpha$~galaxies confirm previous observations that at a given stellar mass, more \hi-rich galaxies typically have higher SFRs \citep{wang11,huang12}. Since a high \hi~gas fraction relative to other galaxies of a similar mass may indicate the presence of recently accreted gas \citep[\eg][]{moran12}, this trend may suggest that \hi~accretion actively fuels star formation in low-redshift galaxies. Nevertheless, the scatter in the \mhi/SFR~ratio at a constant stellar mass shows that additional factors affect the ability of galaxies to access fresh \hi~fuel and efficiently convert it to \htwo. In particular, the starbursts generally show high \hi-based star formation efficiencies (\ie~short \hi~depletion times) compared to similar mass galaxies (\S~\ref{sec_sfe}). We also find that the starbursts' \hi~gas fractions are not unusually elevated for their stellar mass, despite their much higher SFRs (\S~\ref{sec:hisbs}). Therefore, in most cases, a high efficiency of \hi~to \htwo~conversion, rather than a large influx of gas, is generating their high levels of star formation. Dynamical disturbances are a likely cause of this enhanced efficiency, allowing the starbursts to transport \hi~gas inward and generate high gas column densities more easily than non-starbursts \citep[\eg][]{barnes91,dimatteo07,hopkins13a}.

The balance between \hi\ infall and feedback may lead to a relatively stable \hi~supply in gas-rich starbursts. Addition of \hi~gas through interaction with a gas cloud or gas-rich galaxy may initially raise the \hi~gas fraction. This excess \hi~quickly disappears, however, as gas compression or turbulence resulting from these interactions efficiently converts the \hi~gas to \htwo~and stars. As the starburst progresses, the \hi~gas fraction may stabilize instead of continuing its decrease. The starburst may not affect \hi~gas in the outer galaxy, and photodissociation of molecular gas may replenish \hi~in the inner regions. Indeed, previous observations have demonstrated that photodissociated gas may be a significant component of the ISM in the inner regions of starburst galaxies \citep[\eg][]{stacey91}.  Radiative feedback does not appear sufficient to deplete the \hi~gas in the starbursts through ionization, and the two high EW starbursts with the highest ratios of H$\alpha$~luminosity to \hi~mass (\ie~the shortest $t_{\rm dep}$) also show the highest \hi~gas fractions. After the starburst, the remaining \hi~reservoir may continue to fuel star formation, or additional feedback mechanisms, such as supernovae, may ultimately quench the burst.

\subsection{The Link Between \hi, Metallicity, and Dust}
For the ALFALFA~H$\alpha$~sample as a whole, we find that galaxies' sSFRs and \hi~gas fractions are only weakly correlated (\S~\ref{sec:hisbs}). This result contrasts with the strong observed correlation between galaxy color and \hi~gas fraction identified in previous studies \citep[\eg][]{kannappan04,zhang09,catinella10,huang12} which suggests that galaxy sSFRs are linked to their \hi~content \citep[\eg][]{zhang09,huang12}. We likewise observe a tight correlation between \hi~gas fraction and $g-r$~color. For our sample of gas-rich galaxies, we demonstrate that dust extinction, rather than star formation, drives the tight trend between \hi~and color. 

In essence, the \hi~and color relation is a manifestation of the well-known galaxy mass-metallicity relation \citep[\eg][]{lequeux79,tremonti04}. Recent studies suggest that galaxies in fact lie on a fundamental plane of stellar mass, SFR, and metallicity \citep{laralopez10,mannucci10}, known as the `fundamental metallicity relation'. Gas content likely drives this relation \citep[\eg][]{dave12,lilly13}, and \citet{bothwell13}~show that a fundamental relation exists between \hi~mass, stellar mass, and metallicity. In this model, \mhi/$M_*$~and metallicity should anti-correlate. Since more metal-rich galaxies have higher dust content, we would therefore expect galaxies with high \mhi/$M_*$~to have less dust extinction and bluer colors. 

Several processes could produce the proposed relation between galaxy stellar mass, \hi~content, and metallicity. Over their lifetimes, galaxies with higher stellar masses will have both consumed more \hi~gas and produced more metals via star formation. In addition, the higher potential wells of more massive galaxies should allow them to retain metals more effectively \citep[\eg][]{dave11b}. At a given stellar mass, galaxies that have experienced a recent inflow of metal-poor gas from the IGM should have lower overall metallicities as well as higher \hi~content. Finally, we note that the elevated dust content of metal-rich galaxies will also allow them to convert their \hi~gas to H$_2$~more efficiently by shielding molecular gas from UV radiation \citep[\eg][]{krumholz09, bolatto11}, thereby reducing their \hi~fractions further. This connection between \hi~content and dust extinction provides a natural explanation for the observed relation between \hi~gas fraction and color. \hi~inflows and outflows are fundamentally important to the evolution of galaxy metallicities, while galaxy metallicity may also influence the \hi~supply by promoting H$_2$~formation.

\subsection{Galaxy Structure and \hi~Conversion}
Another factor that may enhance the conversion of \hi\ to H$_2$~is galaxy structure. We find that the \hi~depletion time in disk-dominated systems anti-correlates with stellar surface density (\S~\ref{sec_sfe}). This relation may indicate that higher disk mid-plane pressure aids the formation of molecular clouds and increases the H$_2$/\hi~ratio, as proposed by \citet{blitz06}. However, alternate theories of H$_2$~formation based on self-shielding \citep[\eg][]{krumholz09} may result in a similar stellar surface density scaling relation \citep{fu10}. Early supernova feedback in low-mass galaxies may lead to the delayed infall of \hi~gas, resulting in an increased \hi~content and longer depletion time at the present day \citep{fu10}. By suppressing past star formation episodes, this efficient feedback should also lead to lower present-day stellar surface densities. 

We only observe this correlation between \hi~depletion time and stellar surface density in disk galaxies, however. We show that high-mass, early-type galaxies from the GASS sample \citep{catinella13} do not follow a similar relation. Within this early-type sample, we find no correlation between \hi~depletion time and morphology, which suggests that bulge strength is not affecting the current SFE in these galaxies. Instead, we suggest that the scatter in SFE results from differences in the spatial distribution and surface density of the \hi~gas. 

\subsection{Mergers and \hi~Conversion}
For several of the starbursts, major mergers may induce their high SFRs and high \hi~to \htwo~conversion efficiencies (\S~\ref{sec_merge}). During interactions, tidal torques propel gas toward the center of galaxies, thereby raising the gas column density \citep[\eg][]{barnes91} and allowing galaxies to access their external \hi~supply. Furthermore, turbulent motions during the merger may compress gas clouds \citep{elmegreen93}, leading to a higher fraction of dense gas and a higher H$_2$/\hi~ratio \citep{powell13}. Starbursts also appear to form stars from H$_2$~more efficiently than non-starbursts \citep[\eg][]{kennicutt98}, perhaps due to the higher mean density and shorter free-fall time in their ISM \citep{krumholz12b}. A merger scenario could therefore explain the moderate \hi~but high SFRs of at least some of the ALFALFA~H$\alpha$~starbursts.

Evidence of morphological and kinematical disturbances support a merger origin for several of the starbursts. We find that the high EW starbursts' optical morphologies are more asymmetric and their \hi~velocity profiles have wider wings than most of the non-starbursts. Merger simulations predict that the peak star formation activity should occur near coalescence \citep[\eg][]{mihos94,cox08,lotz10a}, consistent with the starbursts' morphologies. The highest asymmetries, however, should appear shortly before coalescence. Interestingly, our two most asymmetric starbursts, which appear to be in this merger stage, also have higher-than-average \hi~gas fractions for their stellar mass. Their high \hi~gas fractions suggest that \hi~conversion to \htwo\ may be an ongoing process during merger-driven star formation \citep[\eg][]{hibbard96}. 

Due to the unresolved nature of the Arecibo \hi~observations, we do not know the surface density of the \hi~gas in the ALFALFA~H$\alpha$~starbursts. However, we find that they have above-average ratios of \mhi~to the galaxy optical area. Therefore, either they lie closer to the maximum column density threshold for \hi, which would aid H$_2$~formation, or their \hi~is significantly more extended than the optical disk. For instance, extended \hi~tidal tails may contain a large fraction of the \hi~gas in merging galaxies \citep[\eg][]{hibbard96}. 

The evidence for mergers among the lowest mass starbursts is less clear, in part due to their lower S/N optical images and \hi~spectra. Their \hi~velocity profiles do not noticeably differ from the profiles of low-mass non-starbursts, which may indicate a lack of kinematical disturbances. Resolved \hi~observations likewise indicate that ordered kinematics are not uncommon in dwarf starbursts \citep{lelli14}. However, if the dwarf starbursts experience periodic bursts, any initial disturbance may no longer be evident \citep[\eg][]{verbeke14}. We observe an increased scatter in SFE at the low-mass end of our sample, which supports this scenario of episodic bursts in dwarf galaxies. Supernova feedback may have a stronger effect in the low potential wells of dwarf galaxies, leading to temporary quenching and a renewed burst of star formation as neutral gas falls back into the galaxy \citep[\eg][]{lee07,verbeke14}. Higher resolution imaging and \hi~spectral observations will be necessary to determine whether or not the dwarf starbursts are merging systems.

Finally, the spiral structure and orderly \hi\ velocity profiles of the three highest mass starbursts, AGC 330186, UGC 470, and UGC 12821, suggest they are not experiencing a major merger. UGC 470 has a higher than average gas fraction for its stellar mass and is known to have an unusually extended \hi\ disk \citep{dowell10}. It is also the only starburst in the sample with a typical \hi\ depletion time for its stellar mass or $\Sigma_{\rm SFR}$ (Figures~\ref{fig_gtbins}~and \ref{fig_ssd_gt}). Unlike the other starbursts, UGC 470's excess gas supply, rather than a high \hi-to-\htwo\ conversion efficiency, may account for its enhanced star formation. Minor mergers or other minor disturbances may temporarily increase the SFR for AGC 330186 and UGC 12821, and most of the \hi\ disk may remain undisturbed.

\subsection{\hi~Cycles in Starbursts}
We therefore propose the following picture for the cycling of \hi~and star formation throughout the strongest starburst episodes. In the intermediate mass starbursts, a gas-rich major merger triggers the star formation. As the two galaxies approach coalescence, the total stellar mass of the system increases. Since the two individual galaxies have \hi~gas fractions typical of lower mass galaxies, the combined system appears to have a higher \hi~mass than other galaxies of a similar total mass. Strong tidal torques drive \hi~gas outward to form tidal tails and inward to fuel star formation. At this time, the system  appears progressively more morphologically disturbed, and the increased gas flows and turbulence create kinematically disturbed \hi~profiles. The two most asymmetric starbursts in our sample, AGC 330517 and AGC 330500, are possible examples of this stage, exhibiting disturbed morphologies, disturbed \hi\ kinematics, and elevated \hi\ gas fractions. As the \hi~gas flows inward, turbulence and gas compression efficiently convert \hi~to H$_2$, reducing the \hi\ gas fraction and increasing the SFR. Although the gas flows bring \hi~inward, star formation is not necessarily restricted to the nuclear region, consistent with recent simulations \citep[\eg][]{powell13} and with the varied H$\alpha$\ morphologies of the ALFALFA~H$\alpha$\ starbursts (\S~\ref{sec_merge}). Turbulent, dense clumps may arise throughout the merging disks, and the precise merger configuration affects the spatial distribution of star formation. The influx of \hi~gas raises the ISM column density, and the star formation rate increases to the point where supernova-driven turbulence supports the enhanced weight of the ISM \citep{ostriker11}. 

Peak star formation occurs near final coalescence, as the morphological disturbances are fading \citep[\eg][]{lotz10a}, which may explain the slightly lower asymmetries of some of the starbursts (\eg AGC 120193 and AGC 331191). However, at the time of the peak SFR, the turbulent motions driving star formation are still high \citep{powell13} and the \hi~kinematics should still show higher velocities, as we observe in the high velocity wings of the high EW ALFALFA~H$\alpha$\ starbursts' \hi\ profiles. The \hi~content also begins to drop, due to the enhanced H$_2$~fraction and photoionization. Nevertheless, the starbursts still maintain a large \hi~supply, comparable to similar mass non-starburst galaxies (\S~\ref{sec:hisbs}). Radiative feedback does not completely ionize the starbursts' \hi~reservoirs and may only ionize the \hi~gas near the starburst region \citep[\eg][]{hanish10}.  In addition, \htwo~photodissociation may compensate for the ionization of \hi~in the inner regions. Consistent with this scenario, recent \hi\ observations of post-merger galaxies demonstrate that star formation and feedback do not noticeably deplete the \hi\ reservoirs of merging galaxies \citep{ellison15}. Ultimately, gas consumption or feedback terminates the starburst, and the final merged galaxy exhibits a higher stellar mass, higher metallicity, and lower \hi~gas fraction than its individual progenitors.

Lower mass galaxies may experience recurrent starbursts, leading to the enhanced scatter in $t_{\rm dep}$ and sSFR in the low-mass end of our sample. They may accumulate \hi~from the IGM until a dynamical disturbance initiates strong star formation. For instance, this initial trigger could be a merger with another galaxy, an interaction with a dark matter subhalo \citep{helmi12}, or a merger with a gas cloud. In the case of a gas cloud merger, the initial interaction should raise the \hi~content and disturb the \hi~kinematics, but may not trigger an immediate starburst \citep{verbeke14}. Subsequent infall from the gas cloud or enhanced turbulence may then induce a later starburst. The morphology of the star formation should vary depending on the nature of the interaction and the gas cloud trajectory, consistent with the range of H$\alpha$\ morphologies we observe (\S~\ref{sec_merge}). 

Regardless of the cause of the initial burst, supernova feedback rapidly quenches the initial starburst due to the low potential well of the galaxy \citep[\eg][]{dekel86,stinson07,hopkins13}. This feedback may increase the \hi~velocity dispersion, but these kinematic disturbances should lag the peak star formation \citep{verbeke14}. After the starburst dies down, the re-accretion of previously expelled gas may trigger a series of subsequent starbursts \citep[\eg][]{lee07,stinson07}. In addition, massive star clusters formed in a prior burst may generate torques that continue to drive \hi~gas inward \citep[\eg][]{elmegreen12} to fuel future starbursts. Thus, low-mass starbursts may lack the clear signs of kinematic disturbances that should characterize higher-mass interacting starbursts, as illustrated by the similar \hi\ velocity profiles of the low-mass starbursts and non-starbursts in our sample (\S~\ref{sec_merge}). These starburst cycles may gradually reduce a dwarf galaxy's \hi~reservoir but are unlikely to deplete it entirely, given the starbursts' high \hi~gas fractions. As with the more massive starbursts, photoionization may not affect the outer \hi~gas and may be balanced by photodissociation near the starburst. Consequently, the \hi\ gas fractions of low-mass starbursts may also remain relatively constant, with little variation with respect to non-starbursts, and may provide a plentiful neutral gas supply capable of fueling multiple starburst episodes. The similar \hi\ gas fractions of the ALFALFA~H$\alpha$\ starbursts and non-starbursts agree with this scenario (\S~\ref{sec:hisbs}), as do observations of the \hi\ content of interacting dwarf galaxies relative to isolated dwarfs \citep{stierwalt14}. By quickly quenching star formation, the high feedback efficiencies in the dwarf starbursts may prevent rapid increases in metallicity or stellar mass. This scenario is consistent with the flat star formation histories and relatively inefficient star formation inferred for dwarf galaxies \citep[\eg][]{behroozi13}. 

The dwarf starbursts' periodic star formation mode may be particularly relevant to high-redshift star formation. Milky Way progenitors at $z\approx2$~may accrete gas from the IGM in discrete episodes that trigger enhanced star formation \citep{woods14}. However, as with the dwarf starbursts, feedback efficiently expels the gas and delays its consumption \citep[\eg][]{woods14}. As a result, these galaxies may have variable SFRs and may maintain an elevated \hi~content that fuels later star formation \citep[\eg][]{hopkins13}. However, unlike $z=0$~galaxies, $z=2$~galaxies experience higher accretion rates from the IGM \citep[\eg][]{keres05}. High-redshift galaxies may therefore have higher average SFRs and larger gas reservoirs than we observe for the ALFALFA~H$\alpha$~sample. Feedback and variable SFRs in low-mass galaxies also have important implications for the reionization of the IGM. \citet{wyithe13}~argue that by suppressing star formation, efficient feedback may reduce the contribution of the lowest mass galaxies to reionization. Constraining the starburst duty cycle in dwarf galaxies is therefore important to understand both galaxy star formation histories and reionization.

\section{Summary}
\label{sec:summary}

The ALFALFA~H$\alpha$~survey represents the first opportunity to compare the properties of gas-rich starbursts and non-starbursts within a statistically uniform \hi-selected sample. In this work, we analyze the \hi~gas fractions, \hi~depletion times, \hi~kinematics, and optical morphologies of 14 starbursts within the 565 galaxies that make up the ALFALFA~H$\alpha$~Fall-sky sample. This sample illuminates the roles of gas accretion and feedback in determining the \hi~content of starburst galaxies and in triggering and sustaining star formation.

Our main results are as follows:

\begin{enumerate}
\item On average, the ALFALFA~H$\alpha$~galaxies with higher instantaneous sSFRs tend to have slightly higher \hi~gas fractions, but this trend is weak and shows substantial scatter. Galaxies with sSFRs that differ by an order of magnitude may still have the same \hi\ gas fraction, and most of the starburst galaxies have \hi\ gas fractions similar to galaxies with significantly lower SFRs. In contrast, we observe a tight trend between \hi~gas fraction and $g-r$~color. We show that dust extinction, rather than recent star formation, is primarily responsible for the tight \hi-color correlation. This link between dust extinction and \hi~gas fraction likely stems from the relation between stellar mass, metallicity, and \hi~content in galaxies \citep{bothwell13}.

\item Disk galaxies lie on a sequence of decreasing \hi~depletion time with increasing stellar surface density. The observed trend is consistent with the idea that higher midplane pressures encourage the formation of \htwo~from \hi~\citep{blitz06}. Disk galaxies from the GASS sample \citep{catinella10,catinella13} also fall on this sequence, while spheroid-dominated systems are offset to higher $t_{\rm dep}$. The spread in the \hi~depletion times of the spheroids reflects a spread in sSFR, but shows no trend with either \hi~gas fraction or bulge-to-disk ratio. Instead, the spatial distribution of \hi~in early-type galaxies likely determines whether gas clouds can reach the necessary densities to form molecular gas.

\item Gas-rich starbursts are able to maintain a relatively constant \hi~supply. Most of the 14 starbursts show little to no increase in \mhi/$M_*$\ or $\Sigma_{\rm HI}$\ relative to galaxies of a similar mass; by the time of the starburst episode, any excess atomic gas has already been converted into \htwo. However, we do find a few exceptions to this scenario. The extended \hi\ disk of UGC 470 may fuel its elevated star formation. In addition, the two most optically asymmetric starbursts, which appear to be in a pre-coalescent merger stage, do show some evidence for enhanced \hi~gas fractions. These asymmetric galaxies suggest that the conversion of excess \hi~to \htwo~is an ongoing process during mergers. 

\item Ionization does not appear to substantially deplete the starbursts' \hi~gas, and photodissociation of \htwo~may compensate for decreases in \hi~due to consumption and ionization. Although we do not find that starbursts are unusually \hi-rich, we also do not find that starbursts are \hi-deficient, as suggested for starbursts in the SINGG sample \citep{oey07}. Instead, high $\Sigma_{\rm SFR}$ galaxies span a wide range of \hi~gas fractions. The similarity of the \hi~gas fractions of starbursts and non-starbursts may indicate that the intense ionizing radiation of the starbursts does not penetrate to the outermost regions hosting much of the \hi~mass. 

\item The starbursts use their \hi\ more efficiently than the rest of the sample, as indicated by their lower \hi~depletion times relative to galaxies of a similar mass or stellar surface density. Major mergers may cause these high efficiencies in at least some of these starbursts, as suggested by the starbursts' asymmetric optical morphologies and the wide \hi~velocity profile wings in several starbursts. The high optical asymmetries of two starbursts are consistent with major mergers. The lower, but slightly elevated asymmetries of an additional six starbursts may indicate either more minor disturbances or mergers near coalescence, the merger phase predicted to cause the largest enhancement in SFR. Finally, consistent with recent merger simulations \citep[\eg][]{powell13}, we find that extended, rather than nuclear, star formation may dominate the morphologies of some starbursts.

\item While some of the starbursts are likely mergers, the lowest mass starbursts, with $M_*<10^8$\Msol, do not show clear evidence of disturbed optical morphologies or \hi~kinematics. These dwarf starbursts may undergo periodic bursts, possibly triggered by a previous interaction, in which case unusual gas kinematics might not be apparent \citep{verbeke14}. Large fluctuations in SFR appear characteristic of dwarf galaxies \citep[\eg][]{lee07}, and the dwarf galaxies in our sample also show a larger scatter in SFE than more massive galaxies. The large, apparently sustainable \hi~gas fractions of low-mass starbursts may provide ample fuel for multiple generations of starbursts.

\end{enumerate}

The ALFALFA~H$\alpha$~galaxies demonstrate that while starbursts may differ dramatically from non-starbursts in their molecular gas content, the atomic gas fractions of starbursts and non-starbursts are similar. Efficient conversion of atomic to molecular gas reduces any \hi~excess, and the localized starburst may not affect the galaxy's extended \hi~reservoir. The \hi~gas fractions of most low-mass starbursts may remain approximately constant throughout the burst, and only the most extreme starbursts at low redshift may significantly disrupt their \hi~gas supply.

\acknowledgments{We thank the anonymous referee for a thoughtful and prompt report. We thank Gus Evrard, Lee Hartmann, Shan Huang, and TJ Cox for helpful discussions and suggestions. We are grateful to Betsey Adams for information on the ALFALFA observations, data processing, and data analysis codes. A.E.J.\ acknowledges support from an NSF Graduate Research Fellowship. A.V.S acknowledges support by fellowships from the Indiana Space Grant Consortium. M.S.O acknowledges support from NSF AST-0806476 and thanks the Cornell Astronomy Department for sabbatical hospitality. The ALFALFA~H$\alpha$\ project was supported by NSF grant AST-0823801, by The College of Arts and Sciences at Indiana University, and by NOAO Survey Programs. We thank the entire team, particularly R. Giovanelli, for their efforts in observing and data processing that produced the ALFALFA source catalog. The ALFALFA team at Cornell is supported by NSF grant AST-1100968 and by the Brinson Foundation.

The Arecibo Observatory is operated by SRI International under a cooperative agreement with the National Science Foundation (AST-1100968), and in alliance with Ana G. M{\'e}ndez-Universidad Metropolitana, and the Universities Space Research Association.  

}


\end{document}